\documentclass[]{emulateapj}
\usepackage{pstricks}

\newcommand{\eqref}[1]{(\ref{#1})}
\newcommand{\myincludegraphics}[2]{\includegraphics[clip,#1]{#2}}
\newcommand{\reftext}[1]{{\relax #1}}

\shorttitle{Chemical evolution of protoplanetary disks}
\shortauthors{Heinzeller et al.}

\begin{document}

\title{Chemical evolution of protoplanetary disks -- the effects of viscous accretion, turbulent mixing and disk winds}

\author{D. Heinzeller\altaffilmark{1} and H. Nomura}
\affil{Department of Astronomy, Graduate School of Science, Kyoto University, Kyoto 606-8502, Japan}
\email{dominikus@kusastro.kyoto-u.ac.jp}

\and

\author{C. Walsh and T. J. Millar}
\affil{Astrophysics Research Centre, School of Mathematics and Physics, Queen's University Belfast,\\ Belfast, BT7 1NN, UK}
\altaffiltext{1}{\reftext{Current address: Meteorological Service of New Zealand Ltd., 30 Salamanca Rd, Kelburn, Wellington 6012, New Zealand}}
\begin{abstract}
We calculate the chemical evolution of protoplanetary disks considering radial viscous accretion,
vertical turbulent mixing and vertical disk winds. We study the effects on the disk chemical structure 
\reftext{when} different models for the formation of molecular hydrogen on dust grains
\reftext{are adopted.}
Our gas-phase chemistry is extracted from the UMIST Database for Astrochemistry (Rate06)
to which we have added detailed gas-grain interactions. 
We use our chemical model results to generate synthetic near- and mid-infrared \reftext{LTE} line emission spectra  
and compare these with recent Spitzer observations. 

Our results \reftext{show} that 
\reftext{if H$_2$ formation on warm grains is taken into consideration, the H$_2$O and OH abundances in the
disk surface increase significantly}.
We find the radial accretion flow strongly influences the molecular abundances,
with those in the cold midplane layers particularly affected. 
On the other hand, we show that diffusive turbulent mixing affects the disk chemistry in the warm molecular
layers, influencing the line emission from the disk and subsequently improving agreement with observations. 
We find that NH$_3$, CH$_3$OH, C$_2$H$_2$ and \reftext{sulphur-containing} species are
\reftext{greatly enhanced}
by the inclusion of turbulent mixing. We demonstrate that disk winds potentially affect the disk chemistry and the resulting 
molecular line emission in a similar manner to that found when mixing is included.
\end{abstract}

\keywords{astrochemistry; protoplanetary disks; accretion, accretion disks; turbulence; infrared: planetary systems}

\section{Introduction}
Protoplanetary disks are a natural and active environment for the creation of simple and complex molecules.
Understanding their physical and chemical evolution is key to a deeper insight into
the processes of star and planetary system formation and corresponding chemical
evolution. Protoplanetary disks are found encompassing young stars in star-forming regions
and typically disperse on timescales of less than a few million years \citep[e.\,g.,][]{haisch01,haisch06,watson07}.
Observations of thermal dust emission from these systems date back to the 1980s
\citep[see, e.\,g.,][]{lada84,beckwith90} and gave important insight into the structure and evolution of
disks \citep[see, e.\,g.,][for a review]{natta07}, 
however, inferring the properties of protoplanetary disks from dust continuum observations has several 
\reftext{shortcomings: the} dust component is only $\approx$ $1\%$ of the total mass contained in the disk and 
its properties are expected to change with dust growth and planet formation. 
Also, dust spectral features are too broad to study the disk kinematics
\citep[see, e.\,g.,][]{carmona10}, and observations suggest that the gas and dust temperatures can differ considerably,
particularly in the UV- and X-ray heated surface layers of the inner disk \citep[e.\,g.,][]{qi06,carmona07}.

\bigskip
The observation of molecular line emission 
from disks at (sub)mm wavelengths is limited by the low spatial resolution
and sensitivity of existing facilities and target the 
rotational transitions of abundant molecules in the outer ($\geq 50\,\mathrm{AU}$) regions
of the disk \citep[see, e.\,g.,][for a review]{dutrey07}. 
Planet formation, however, most likely occurs at much smaller radii \citep{armitage07}.
The superior sensitivity and spatial and spectral resolution of 
the Atacama Large Millimeter Array (ALMA), currently under construction, is 
required to observe the (sub)mm emission lines from
the inner regions of protoplanetary disks in nearby star-forming regions 
to enable direct study of the physical and chemical structure of \reftext{the inner disk}.

Near- and mid-infrared observations, on the other hand, can probe the warm 
planet-forming regions \citep[see, e.\,g.,][for a review]{najita07}.
The $4.7\micron$ CO line emission has been detected at 
radii $<1\,\mathrm{AU}$, using, for example, Keck/NIRSPEC, Gemini/Phoenix and VLT/CRIRES
\citep[e.\,g.,][]{carr07,pontoppidan08,brittain09,vanderplas09}. 
The Spitzer Space Telescope has been used to spectroscopically study the inner disk at mid-infrared wavelengths.
These observations show a rich spectrum of emission lines from simple 
organic molecules \citep[e.\,g.,][]{carr08,salyk08,pontoppidan10} as well as from complex 
polycyclic aromatic hydrocarbons (PAHs) \citep[e.\,g.,][]{habart06,geers06}. 
They suggest high abundances of H$_2$O, HCN, C$_2$H$_2$ and provide evidence that the disk supports an active chemistry.
C$_2$H$_2$ and HCN have also been detected in absorption towards IRS\,46 and GV\,Tauri \citep{lahuis06,gibb07}.
A particularly interesting case is CO$_2$, which has been reported to be abundant in the circumstellar disk of AA\,Tauri at
$\sim 1.2\,\mathrm{AU}$ \citep{carr08}, in IRS\,46 \citep{lahuis06} and in several other sources 
\citep[e.\,g.,][]{pontoppidan10}, but which is not detected in the majority
\reftext{of sources observed by Spitzer}.
The recently launched Herschel Space Observatory and future space telescopes 
(e.\,g., the Space Infrared telescope for Cosmology and Astrophysics
\reftext{(SPICA)} and the James Webb Space Telescope \reftext{(JWST)}) 
will enable the observation of molecular line emission in the mid- to far-infrared 
with unprecedented sensitivity and spectral resolution and will help address these controversial findings.

Due to increasing computational power, the theoretical study of the physical and 
chemical evolution of protoplanetary disks is entering its Golden Age. 
As a consequence of the observational limitations in the pre-Spitzer era,
many numerical models focused only on the outer disk \citep[e.\,g.,][]{aikawa02,willacy07}.
With the launch of Spitzer, however, the chemistry of the inner disk has increased in importance 
\citep[e.\,g.,][]{woods09}.
Unfortunately, complete models, including the physical and chemical evolution of the gas and the dust in the disk, 
coupled with a consistent treatment of the radiative transfer, 
is still beyond reach and several compromises have to be made.

One possibility is to ignore the chemical evolution and assume 
chemical equilibrium holds throughout the disk to investigate the physical structure
in detail \citep[e.\,g.,][]{kamp04,jonkheid04}. 
Pioneering work by \citet{gorti04} and \citet{nomura05} combined a basic disk chemistry
with a self-consistent description of the 
density and temperature structure taking into account the gas thermal balance, the stellar irradiation
and the \reftext{dust grain} properties. 
Recently, \citet{woitke09} calculated radiation thermo-chemical
models of protoplanetary disks, including frequency-dependent 2D dust continuum radiative transfer, 
kinetic gas-phase and \reftext{photochemistry}, ice formation and non-LTE heating and cooling. 
These studies show, in accordance with the observations, 
that the stellar UV and X-ray irradiation increases the gas temperature 
considerably in the surface layer of the disk, 
altering the disk chemistry drastically in this region.

Converse to that described above, several authors instead 
focused on combining the physical motion of 
material in the disk with time-dependent chemical network calculations
\citep[e.\,g.,][]{aikawa99,markwick02,ilgner04,nomura09}. 
Radial inward motion due to the viscous accretion and vertical motion due
to inherent turbulent mixing or disk winds have the potential to change the disk chemistry 
significantly, depending on the timescales of
the transport processes. Molecules, initially frozen out on dust grains, can evaporate 
and increase the gas-phase abundances,
provided that accretion and vertical mixing is comparable
to, or faster than, the chemical reactions and the destruction by incident stellar irradiation. 
First steps towards including the accretion flow in the chemical network 
have been carried out by \citet{aikawa99},
\citet{markwick02} and \citet{ilgner04}. \citet{nomura09} calculated the effect of accretion along 
streamlines in steady $1$+$1$-dimensional
$\alpha$-disk models for chemical networks containing more than 200 species.
They found that the chemical evolution is influenced globally by the inward accretion process, 
resulting in, e.\,g., enhanced methanol
abundances in the inner disk. \citet{ilgner04} also studied vertical mixing using the diffusion 
approximation and concluded that vertical
mixing changes the abundances of the sulphur-bearing species. Their disk model, however, assumed 
the gas and dust temperatures were coupled everywhere in the disk and ignored
stellar irradiation processes. \citet{semenov06} investigated gas-phase 
CO abundances in a 2D disk model, allowing for
radial and vertical mixing using a reduced chemical network of 250 species and 1150 chemical 
reactions, and find the mixing processes strongly affect 
the CO abundances in the disk. \citet{willacy06} studied 
the effects of vertical diffusive mixing driven by turbulence
and found that this can greatly affect the column density of various 
molecules and increase the size of the intermediate, warm molecular layer in the disk.

In this study, we consider the effects of inward accretion motion and 
vertical turbulent mixing to investigate to what extent the abundances of molecules
such as H$_2$O, CO$_2$ and CH$_3$OH are affected in the inner disk. 
As an alternative driving force for vertical transport,
we investigate the effect of upward disk winds on the disk chemistry. Several kinds of disk
wind models, such as magnetocentrifugally driven winds \citep[e.\,g.,][]{blandford82,shu94}
and photoevaporation winds \citep[e.\,g.,][]{shu93} have been suggested thus far. Recently,
\citet{suzuki09} studied disk winds driven by MHD turbulence in the inner disk and found
that large-scale channel flows develop most effectively at $1.5$--$2$ disk scale heights, 
i.\,e., approximately in those layers where molecular
line emission is generated. The disk model in our study is from \citet{nomura05} including X-ray heating as described in 
\citet{nomura07} and accounts for decoupling of the gas and dust temperatures in the disk surface due to
stellar irradiation. 
We use a large chemical network and allow for adsorption and desorption onto and from dust grains.
We also calculate mid-IR dust continuum and molecular line emission spectra assuming local thermodynamic equilibrium (LTE) 
and compare our results to recent observations. 
In Section~\ref{sec_model} we describe our model in detail. 
In Section~\ref{sec_results}, we present and discuss
our results, before drawing our final conclusions in Section~\ref{sec_conclusion}.

\section{Model}\label{sec_model}
\subsection{Disk structure}
Our physical disk model is from \citet{nomura05} with the addition of X-ray heating 
as described in \citet{nomura07}. They compute the physical structure of a steady, axisymmetric protoplanetary disk 
surrounding a typical T\,Tauri star with mass $M=0.5M_\odot$, radius $R=2R_\odot$,
effective temperature $T_\mathrm{eff}=4000\,\mathrm{K}$ and luminosity $L=0.87L_\odot$.
We adopt the $\alpha$-disk model \citep{shakura73} to obtain the surface density profile
\reftext{from an equation for angular momentum conservation \citep[e.\,g.,][Eq.~(3.9)]{pringle81}.}
\reftext{We assume an accretion rate of $\dot{M}=10^{-8}M_\odot\mathrm{yr}^{-1}$ and a viscosity parameter of $\alpha=0.01$.}

We denote the radius in the cylindrical coordinate system with $s$ to distinguish it from the radial 
distance from the central star, i.\,e., $r = \sqrt{s^2 + z^2}$, where $z$ represents the height from the disk 
midplane.

The disk is calculated between $s=0.01\,\mathrm{AU}$ and $300\,\mathrm{AU}$ in a $1+1$-dimensional model,
\reftext{which corresponds to a total disk mass of $2.4 \cdot 10^{-2} M_\odot$ for the  
values of $\alpha$ and $\dot{M}$ given above}.
The steady gas temperature and density distributions of the disk are obtained self-consistently by iteratively solving the equations for
hydrostatic equilibrium in the vertical direction and local thermal balance between heating and cooling of gas.
To solve the equation for hydrostatic equilibrium 
\reftext{\citep[][Equation~(11)]{nomura07}},  the condition
\reftext{\begin{equation}
\int_{-z_\infty}^{+z_\infty} \rho_\mathrm{gas}(s,z)\,dz = \Sigma_\mathrm{gas}(s)
\end{equation}
with $ \rho_\mathrm{gas}(s,z_\infty) = 10^{-18}\mathrm{g}\,\mathrm{cm}^{-3}$} is imposed \reftext{\citep{nomura05}}.
The heating sources of the gas are grain photoelectric heating by FUV photons and X-ray heating
by hydrogen ionization, while cooling occurs through gas-grain collisions and by line transitions
\reftext{(for details, see \citealt{nomura05}, Appendix~A and \citealt{nomura07}).}

The dust temperature is computed assuming radiative equilibrium between the absorption of stellar radiation
and resulting thermal emission by the dust grains. 
\reftext{Viscous heating
is only taken into account in the disk midplane since its effects in the disk surface are comparably small
for $\alpha=0.01$ \citep{glassgold04}. In the midplane, the density is high and the gas and dust temperature
are tightly coupled. Note that it is often assumed that the gas temperature is coupled with the dust
temperature everywhere in the disk. This is not true in the disk surface which is strongly irradiated by the parent star.
The dust temperature is calculated using an iterative radiative transfer \reftext{calculation}
by means of the short
characteristic method in spherical coordinates, with initial dust temperature profiles obtained from the
variable Eddington factor method \citep[see][for details]{nomura02,dullemond00,dullemond02}.}

The stellar X-ray and UV radiation fields are based on spectral observations of TW\,Hydrae, a classical T\,Tauri star.
The X-ray spectrum is obtained by fitting a two-temperature thin thermal plasma model to archival XMM-Newton data
(mekal model, $L_\mathrm{X}\sim 10^{30}\mathrm{erg}\,\mathrm{s}^{-1}$ for $0.1\,\mathrm{keV}\ldots 10\,\mathrm{keV}$) 
\reftext{and is slightly softer than that of typical T\,Tauri stars with temperatures of about $2\,\mathrm{keV}$ in their
quiescent phase \citep[c.\,f.,][]{wolk05}.} The \reftext{stellar} UV radiation field has 
three components: photospheric blackbody radiation, 
hydrogenic thermal bremsstrahlung radiation and strong $\mathrm{Ly}\,\alpha$ emission
($L_\mathrm{FUV}\sim 10^{31}\mathrm{erg}\,\mathrm{s}^{-1}$ for $6\,\mathrm{eV} \ldots 13\,\mathrm{eV}$).
\reftext{The photochemistry used in this paper is described in detail in \citet{walsh10}.
Note that the scattering of $\mathrm{Ly}\,\alpha$ emission by atomic hydrogen is not taken into account.
On the other hand, the $\mathrm{Ly}\,\alpha$ emission does not affect the molecular abundances significantly
when the dust grains are well mixed with the gas \citep{fogel11}.}
\reftext{We include the interstellar UV radiation field (\citet{draine78} and \citet{vandishoeck82}), 
however, the} interstellar \reftext{UV and X-ray radiation fields are} negligible compared to the central star 
\citep[for details, see][]{nomura07}.

\begin{figure*}
\myincludegraphics{width=0.99\textwidth}{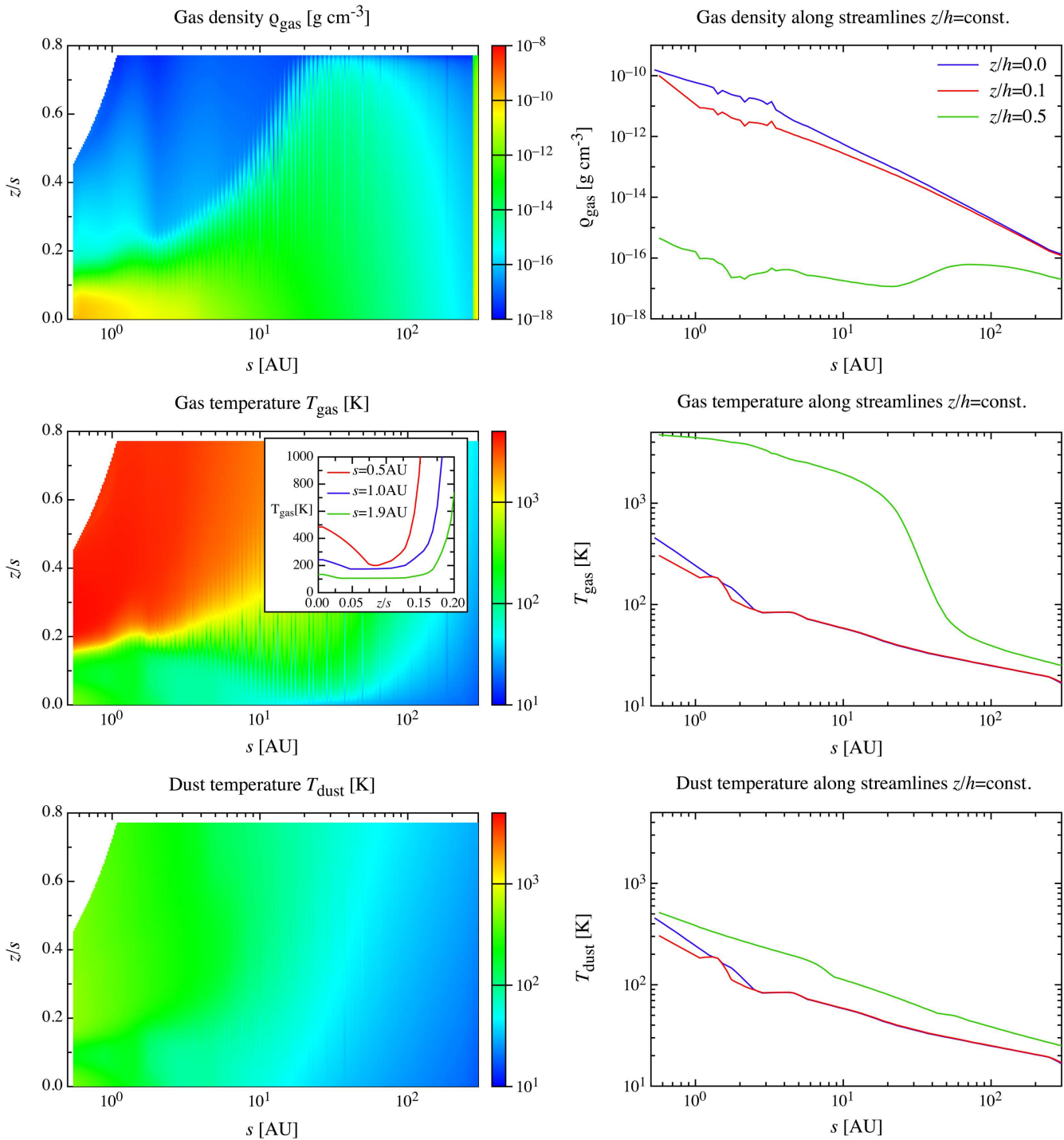}
\caption{From top to bottom: Gas {density}, gas {temperature} and dust {temperature} 
for the full disk {model} (left) and for constant streamlines $z/h$ of the viscous accretion (right),
where $h$ denotes the scale height of the disk at radius $s$. For the gas {temperature}, {inlaid are}
the vertical profiles as function of $z/s$ in the inner disk region.\label{fig_disk_properties}}
\end{figure*}

The mass density, gas temperature and dust temperature for our disk model are 
displayed in Figure~\ref{fig_disk_properties}\reftext{,} top, middle and bottom, respectively. 
A logarithmic spacing of the grid cells is adopted in radial direction,
while the vertical structure is calculated on a linear grid. The number of grid cells used in this
work is $35\times 30$ for $0.5\,\mathrm{AU}\leq s \leq 300\,\mathrm{AU}$ and $0\leq z/s\leq 0.8$.
\reftext{While this resolution is relatively coarse, its consequences for the line emission are
limited due to the interpolation method during the line of sight integration (c.\,f., Section~\ref{sec_line_spectra}).}

A \reftext{more} detailed description of the background theory and calculation of the disk physical model, 
along with further references, is given in \citet{nomura05}\reftext{, \citet{nomura07} and \citet{walsh10}}.
\subsection{Dust properties}
The dust properties are important as they influence the physical and chemical structure
of the disk in several ways. As the dominant source of opacity, they determine the dust temperature
and the UV radiation field in the disk. They also affect the gas temperature, because photoelectric heating by
UV photons is the dominant heating mechanism towards the disk surface. Further, the total
surface area of the dust grains influences the molecular hydrogen abundance through the H$_2$ formation 
rates on grains and determines the freeze-out rate of chemical species onto the grain mantle throughout the disk 
(see Section~\ref{sec_gas_grain_interaction}). 
Finally, due to their dominant contribution to the opacity, the dust grain model also 
determines the emerging continuum spectrum.

For consistency, we adopt the same dust grain model for the spectral calculations as for the underlying protoplanetary
disk \citep{nomura05}. We assume the dust grains to be compact and spherical, and to consist
of silicate grains, carbonaceous grains and water ice. We further assume that the dust and the
gas are well mixed and that the gas-to-dust mass ratio is uniform ($\rho_\mathrm{dust}/\rho_\mathrm{gas}=0.01$)
in the disk. The size distribution is taken from \reftext{\citet{weingartner01a}} to reproduce
the observed extinction in dense clouds\reftext{, with minimum and maximum sizes for the grains
of $3.5$\,\AA\ and \reftext{$\sim$}$10\,\micron$, respectively. The heating of PAHs is included in the model for
photoelectric heating \reftext{\citep{weingartner01b}}. For further details, see also Appendix~D in \citet{nomura05}.}

\subsection{Gas-phase reaction network}
For the calculation of the chemical evolution in the disk, we use a subset of the UMIST Database for
Astrochemistry, or Rate06 \reftext{\citep{woodall07}}, 
neglecting only those species, and thus reactions, containing fluorine and 
phosphorus \citep{walsh10}.
Since we are primarily interested in those regions where molecular line emission originates and where
most molecules are in the gas phase, we neglect grain-surface reactions 
excepting H$_2$ grain formation (see Section~\ref{sec_gas_grain_interaction}).
In summary, our gas-phase reaction network contains
$375$ atomic, molecular and ionic species, ranging from atomic hydrogen H to relatively complex 
molecules such as H$_3$C$_9$N$^{+}$ with
mass $m=125\,\mathrm{u}$ (see Table~\ref{tab_umist_binding_energies} for a list of species of particular
interest in this study and their binding energies on dust grains).
The reaction network consists of $4346$ reactions including $3957$ two-body 
reactions, $214$ photochemical reactions, $164$ X-ray/cosmic-ray induced photoreactions 
and $11$ reactions for direct cosmic-ray ionization. 

\subsection{Gas-grain {interactions}}\label{sec_gas_grain_interaction}
The formation of molecular hydrogen on the surface of dust grains is thought to be the most efficient means of 
forming H$_2$ directly from atomic hydrogen. 
Except for this process, we neglect grain-surface reactions of \reftext{icy mantle} species, but allow
for adsorption onto and desorption from dust grains. Hence, except for atomic hydrogen, 
a particle that hits a dust grain can either detach without change or 
freeze out and remain \reftext{in the icy grain mantle} from where it may eventually 
evaporate. Atomic hydrogen forms
H$_2$, freezes out onto the grain mantle or evaporates without reaction.

For the H$_2$ grain formation rates, we consider two different models. 
In model A, we refer to early work by \citet{gould63} and \citet{hollenbach71},
who derive a grain formation rate for imperfectly shaped grain surfaces of
\reftext{\begin{eqnarray}
\dot{n}(\mathrm{H}_2)^\mathrm{gr}_\mathrm{A}
&=&\frac{1}{2} n_\mathrm{H} \cdot v_\mathrm{H}^\mathrm{th}(T_\mathrm{gas}) n_\mathrm{dust} \pi \langle a^2\rangle \alpha_\mathrm{H}\nonumber\\
&=&n_\mathrm{H} \cdot 2 \cdot 10^{-18}\mathrm{s}^{-1}\ n_\mathrm{tot}\left(\frac{T_\mathrm{gas}}{300\,\mathrm{K}}\right)^{1/2}\!\!\!.
\label{eqn_h2_grain_formation_model_A}
\end{eqnarray}}%
\reftext{In Equation~\eqref{eqn_h2_grain_formation_model_A}, $v_\mathrm{H}^\mathrm{th}(T_\mathrm{gas})$ is the thermal
relative velocity of hydrogen atoms,
\begin{equation}
v_\mathrm{H}^\mathrm{th}(T_\mathrm{gas}) = \left(\frac{8 k_\mathrm{B}T_\mathrm{gas}}{\pi m_\mathrm{H}}\right)^{1/2}\!,
\end{equation}
and $n_\mathrm{dust} \pi \langle a^2\rangle$ the cross section of the dust grains per unit volume, with
$n_\mathrm{dust}\sim 6 \cdot 10^{-10} n_\mathrm{tot}$ and $\langle a^2\rangle = 4.84 \cdot 10^{-12}\mathrm{cm}^2$
for the dust grain model adopted here. Further,  $n_\mathrm{tot}$ and $n_\mathrm{H}$ denote the total number density
of hydrogen nuclei and the number density of atomic hydrogen, respectively.}
The total number density of hydrogen nuclei is calculated from the 
mass density\reftext{, $\rho_\mathrm{gas}$, using}
\[
n_\mathrm{tot} = \frac{\rho_\mathrm{gas}}{(1 + 4y_\mathrm{He}) m_\mathrm{H}} = \frac{\rho_\mathrm{gas}}{1.4 m_\mathrm{H}}\,,
\]
where $y_\mathrm{He}=0.1$ is the fractional elemental abundance of helium. 
Here, we neglect the mass contained in other, less abundant elements.

Alternatively, in model B, the H$_2$ grain formation is calculated according to \citet{cazaux02},
\reftext{\begin{eqnarray}
\dot{n}(\mathrm{H}_2)^\mathrm{gr}_\mathrm{B}&=& \dot{n}(\mathrm{H}_2)^\mathrm{gr}_\mathrm{A} \cdot \varepsilon(T_\mathrm{dust})\nonumber\\
&=&\frac{1}{2} n_\mathrm{H} \cdot v_\mathrm{H}^\mathrm{th}(T_\mathrm{gas}) n_\mathrm{dust} 
	\pi \langle a^2\rangle \alpha_\mathrm{H} \varepsilon(T_\mathrm{dust})\,,
\label{eqn_h2_grain_formation_model_B}
\end{eqnarray}}%
with several updates for the efficiency $\varepsilon(T_\mathrm{dust})$ (Woitke 2009, private communication).
The sticking coefficient, $\alpha_\mathrm{H}$, is taken from \citet{hollenbach79},
and approximated to first order, i.\,e.,
\begin{equation}
\alpha_\mathrm{H} \approx \frac{1}{1 + 0.4 \sqrt{(T_\mathrm{gas}+T_\mathrm{dust})/100\,\mathrm{K})}}\,.
\label{eqn_sticking_coefficient_hydrogen}
\end{equation}

The model for adsorption of gaseous species onto dust grains and desorption of ice species from dust grains is taken
from \citet{woitke09}, who consider collisional adsorption together with thermal desorption, cosmic-ray
induced desorption and UV photodesorption.

We calculate the ice formation rates for all neutral and negatively charged species $k$
in our reaction network,
\begin{eqnarray}
\dot{n}_{k\,\mathrm{ice}}&=&n_k R_k^\mathrm{ads}\label{eqn_ice_rates}\\
&&- n_{k\,\mathrm{ice}}^\mathrm{desorb} \cdot \left( R_k^\mathrm{des,\,th} + R_k^\mathrm{des,\,cr} + R_k^\mathrm{des,\,ph}\right)\,,\nonumber
\end{eqnarray}
where $n_{k\,\mathrm{ice}}$ denotes the number density of ice species $k$ and $n_{k\,\mathrm{ice}}^\mathrm{desorb}$
its fraction located in the uppermost surface layers of the ice mantle. 
This fraction $n_{k\,\mathrm{ice}}^\mathrm{desorb}$ is given by \citep{aikawa96}
\begin{equation}
n_{k\,\mathrm{ice}}^\mathrm{desorb} = \left\{
\begin{array}{ll}
\displaystyle n_{k\,\mathrm{ice}} \quad &\displaystyle  n_\mathrm{ice} < n_\mathrm{act}\\[1.5ex]
\displaystyle n_\mathrm{act}\frac{n_{k\,\mathrm{ice}}}{n_\mathrm{ice}} \quad &\displaystyle  n_\mathrm{ice} \geq n_\mathrm{act}\,,
\end{array}\right.
\end{equation}
with $n_\mathrm{act}=3\cdot 10^{15} \cdot 4\pi\langle a^2\rangle n_\mathrm{dust}$ being the number 
density of active spots in the ice mantle,
and $n_\mathrm{ice}$ being the number density of all ice species ($n_\mathrm{tot} = n_\mathrm{gas} + n_\mathrm{ice}$).
In the following, we summarize the expressions for the rate coefficients and we refer 
readers to \citet{woitke09} for a detailed discussion.

The adsorption rate for species $k$ with mass $m_k$
and thermal velocity $v_k^\mathrm{th}$ is
\begin{equation}
R_k^\mathrm{ads} = \pi \langle a^2\rangle n_\mathrm{dust} \alpha_k v_k^\mathrm{th}\,,
\label{eqn_adsorption_rate}
\end{equation}
where we assume a sticking coefficient of unity for all species \reftext{excepting} hydrogen 
(c.\,f., $\alpha_\mathrm{H}$ in Equation~\eqref{eqn_sticking_coefficient_hydrogen}).
The desorption rates are
\begin{eqnarray}
R_k^\mathrm{des,\,th} &=& \nu_k^\mathrm{osc} \exp\left(-\frac{E_k^\mathrm{bind}}{T_\mathrm{dust}}\right)\,,
\phantom{\int_A^B}\label{eqn_thermal_desorption_rate}\\
R_k^\mathrm{des,\,cr} &=& f_{70\,\mathrm{K}} R_{k, 70\,\mathrm{K}}^\mathrm{des,\,th}\frac{\zeta_\mathrm{CR}}{5\cdot 10^{-17}\mathrm{s}^{-1}}\,,
\phantom{\int_A^B}\label{eqn_cosmic_desorption_rate}\\
R_k^\mathrm{des,\,ph} &=& \pi \langle a^2\rangle \frac{n_\mathrm{dust}}{n_\mathrm{act}} Y_k F_\mathrm{FUV}\,.
\phantom{\int_A^B}\label{eqn_photon_desorption_rate}
\end{eqnarray}
The binding energies $E_k^\mathrm{bind}\,[\mathrm{K}]$,
for the most important species are listed in Table~\ref{tab_umist_binding_energies}.
The vibrational frequency of species $k$ in the surface potential well is
\begin{equation}
\nu_k^\mathrm{osc}=\left(\frac{3\cdot 10^{15} k_\mathrm{B} E_k^\mathrm{bind}}{\pi^2 m_k}\right)^{1/2}\,.
\end{equation}
The expression for cosmic-ray induced desorption is adopted from \citet{hasegawa93}, 
who evaluate the rate coefficient at a temperature of $70\,\mathrm{K}$
with a duty cycle of the grains of $f_{70}=3.16\cdot 10^{-19}$. 
The photo-desorption yields $Y_k$ are simply set to $10^{-3}$ per UV photon \reftext{\citep{westley95}}.
Finally, the cosmic ray ionization rate $\zeta_\mathrm{CR}$ and the FUV field 
$F_\mathrm{FUV}$ in units $[\mathrm{photons}/\mathrm{cm}^{2}/\mathrm{s}]$
are calculated \reftext{from the density profile and the dust opacity adopted in} the disk model \citep{nomura07}.
\reftext{The calculation of the FUV field is based on \citet[][Equations~(6) and~(7)]{nomura05} and accounts for dust scattering.}
\begin{deluxetable}{lrl}
\tablecaption{Molecular binding energies $E^\mathrm{bind}[\mathrm{K}]$\label{tab_umist_binding_energies}}
\tablewidth{0.99\columnwidth}
\tablehead{\colhead{species} & \colhead{$E^\mathrm{bind}[\mathrm{K}]$} & \colhead{reference}}
\startdata
H          & 350  & (5)\\
H$_2$      & 450  & (5)\\
CO$_2$     & 2690 & (3)\\
H$_2$O     & 4820 & (4)\\
CO         & 960  & (1)\\
OH         & 1260 & (5)\\
CH$_4$     & 1080 & (2)\\
NH$_3$     & 3080 & (4)\\
HCN        & 4170 & (2)\\
CH$_3$OH   & 2140 & (5)\\
C$_2$H$_2$ & 2400 & (2)\\
SO$_2$     & 3070 & (5)\\
H$_2$S     & 1800 & (5)
\enddata
\tablerefs{(1) \citet{sandford88}, (2) \citet{yamamoto83}, (3) \citet{sandford90},
(4) \citet{sandford93}, (5) \citet{hasegawa93}}
\end{deluxetable}
\subsection{Initial conditions}\label{sec_initial_conditions}
The initial abundances we use as input are 
listed in Table~\ref{tab_umist_initial_abundances} and we calculate the
chemical evolution over $10^6\,\mathrm{yr}$, the typical lifetime of protoplanetary disks. The initial
abundances for each species $n_{k, k\,\mathrm{ice}, 0}$ include both the ice and the gas component.

We compare the freeze-out timescale $\tau_k^{\mathrm{ads}}$
and evaporation timescale $\tau_k^{\mathrm{des}}$
for each species in a simple model to 
determine whether it exists initially as gas or ice only:
\begin{equation}
\tau_k^{\mathrm{ads}} = \left(R_k^\mathrm{ads}\right)^{-1}\!\!,\qquad
\tau_k^{\mathrm{des}} = \left(R_k^\mathrm{des}\right)^{-1}\!\!.
\label{eqn_adsorption_desorption_timescale}
\end{equation}
For $\tau_k^{\mathrm{ads}}<\tau_k^{\mathrm{des}}$, the species is initially frozen out, 
\reftext{$n_{k\, \mathrm{ice}} = n_{k,0}$} and $n_{k} = 0$. Otherwise,
it exists in the gas phase only, \reftext{$n_{k} = n_{k,0}$} and $n_{k\,\mathrm{ice}} = 0$.

\begin{deluxetable}{lll}
\tablecaption{Initial fractional abundances in the chemical network, including the gas and ice component of each species.\label{tab_umist_initial_abundances}}
\tablewidth{0.6\columnwidth}
\tablehead{\colhead{species} & \colhead{$x_{k, 0}$\tablenotemark{a}} }
\startdata
H          & 2.00e-07\\
H$^{+}$    & 1.00e-10\\
H$_3^{+}$  & 1.00e-08\\
He$^{+}$   & 2.50e-11\\
He         & 1.00e-01\\
CH$_4$     & 2.00e-07\\
NH$_3$     & 1.00e-05\\
H$_2$O     & 1.00e-05\\
C$_2$H$_2$ & 5.00e-07\\
Si         & 3.60e-08\\
N$_2$      & 2.00e-05\\
C$_2$H$_4$ & 1.00e-08\\
CO         & 5.00e-05\\
H$_2$CO    & 4.00e-08\\
CH$_3$OH   & 5.00e-07\\
O$_2$      & 1.00e-06\\
H$_2$S     & 1.00e-06\\
Fe$^{+}$   & 2.40e-08
\enddata
\tablenotetext{a}{The initial fractional abundance is defined as 
\reftext{$x_{k,0} = n_{k,0}/n_\mathrm{tot}$.}}
\end{deluxetable}
\subsection{Physicochemical interaction{s}: viscous accretion, turbulent mixing and disk winds}
Modeling the physicochemical interaction is the key aspect of our study. 
\reftext{Even with} the steady increase in computational power,
this is still a challenging problem which can only be overcome by massive parallel computing 
and by accepting several simplifications.
In the context of a steady disk model, for example, it is a reasonable assumption to 
describe the interaction between the disk physics
and chemistry as a one-way process. 
Hence, we consider the influence of physical processes on the disk chemistry, but neglect
the feedback of the chemistry on the physical disk structure.

Despite this simplification, the calculation of the disk's chemical evolution with interaction is still considerably more 
computationally expensive, the reason being that adjacent grid cells are 
chemically coupled through the radial and vertical exchange of material. 
Accordingly, their evolution must be calculated in parallel with integration time steps that can be shorter than the
the time scale of the chemical reaction network, due to the Courant-Friedrichs-Lewy (CFL) condition \citep{courant28}
of the fluid (see Section~\ref{sec_timescales}). \reftext{More precisely, to compute the models with interaction, we
adopt the following \reftext{procedure: all} grid cells are initialized at the starting time with their initial conditions
(Section~\ref{sec_initial_conditions}) and the transfer coefficients between the grid cells are evaluated (see below
for the calculation of these coefficients for the different transport processes). The chemical evolution of the
\emph{entire disk is calculated in parallel} for time steps that satisfy the CFL condition and the 
transport coefficients are updated \reftext{each iteration}.
This step is repeated until a total age of the disk of $10^6\,\mathrm{yr}$ is reached.}

In the following, we describe each process we consider in this study. 
A detailed discussion of the numerical methods used to compute the chemical evolution 
is given in Appendix~\ref{app_physicochemical_interaction}.

\subsubsection{Radial transport by viscous accretion}
Firstly, we allow for the radial inward transport of material through the
disk {via accretion}. The main parameter describing this effect
is the accretion rate $\dot{M}=10^{-8}M_\odot\mathrm{yr}^{-1}$,
which relates the radial inflow velocity $v_s$ with the 
column density $\Sigma$ at radius $s$,
\begin{equation}
\dot{M} = 2\pi s v_s \Sigma\quad (v_s>0\ \mathrm{for\ radial\ inflow})\label{eqn_continuity_equation_radial}\,.
\end{equation}
We assume that inward accretion occurs along streamlines $l=\sqrt{s^2+z^2}$, 
with $z/h=\mathrm{const}$. Here, $h$ is the scale height of the disk defined
by $h=c_\mathrm{s}/\Omega_\mathrm{K}$, $c_\mathrm{s}$ \reftext{is} the sound speed at the disk midplane
and $\Omega_\mathrm{K}$ is the Keplerian angular frequency.

We denote the net accretion rate of species $k$, i.\,e., 
the net change in its number density, with $\dot{n}_{k,\,\mathrm{acc}}$.
In steady state, the net accretion rate of species $k$ can be written as 
(see Appendix~\ref{app_accretion} for the numerical implementation)
\reftext{\begin{equation}
\dot{n}_{k,\,\mathrm{acc}} = \frac{\partial (\tilde{A} n_k v_l)}{\partial l}\,,\label{eqn_accretion_rate}
\end{equation}
where $\tilde{A}$ is a geometrical factor and corresponds to $\rho_{i+1}/\rho_i$ in \eqref{eqn_accretion_rate_discrete}.}
The boundary conditions for the accretion process at the inner and outer disk 
radii are set as follows: due to the purely inward motion, all material
that flows inwards from the innermost grid cell is removed from the disk, whereas at 
the outer radius $s=300\,\mathrm{AU}$, the accretion timescales become sufficiently long
that we can safely assume $\dot{n}_{k,\,\mathrm{acc}}=0$ for all species. 
In Section~\ref{sec_timescales}, we compare the timescales for viscous accretion
with the chemical timescale and also with corresponding values for the other physical processes considered here.

\subsubsection{Vertical transport by turbulent mixing}
Although the mechanism is not yet fully clear, it is commonly understood
that the source of viscosity in astrophysical disks is turbulence.
Hence, a sufficiently large viscosity, as is the case in our model ($\alpha=0.01$) 
implies that the effect of turbulent mixing is strong, particularly in the inner disk region. 
While this turbulent mixing\reftext{, in principle}, occurs in the radial and vertical directions, 
we ignore the radial turbulent motion and focus on the vertical mass fluxes only.
Furthermore, we assume that the turbulent velocity is independent of the distance $z$ 
from the midplane and thus only a function of radius.

We calculate the mixing in a diffusion-type approach. The diffusion coefficient is set as $K = v_z\,dz$, where
\begin{equation}
v_z = v_{z,\mathrm{mix}} = \alpha' c_\mathrm{s}\,,
\end{equation}
in analogy to the expression for the turbulent velocity in a \reftext{Shakura-Sunyaev} disk model, 
where the turbulent viscosity is

\begin{equation}
\nu_\mathrm{turb} = v_\mathrm{turb} l_\mathrm{turb} = \alpha c_\mathrm{s} h\,.\label{eqn_viscosity_radial_and_vertical}
\end{equation}
We assume $\alpha' = 10^{-3}$ to account for the turbulence in vertical direction only.
For our numerical model, the length scale $dz$ corresponds to the vertical extent of the
grid cells $\Delta z \approx h/30$. The mixing timescales are discussed in Section~\ref{sec_timescales}.

The usual diffusion equation for species \emph{k} \citep[c.\,f., Equation~(3) in][]{willacy06} reduces to
\begin{equation}
\dot{n}_{k,\,\mathrm{mix}} = \frac{\partial}{\partial z}\left( K n_\mathrm{tot} \frac{\partial(n_k/n_\mathrm{tot})}{\partial z}\right)\,.
\label{eqn_mixing_rate}
\end{equation}
In Appendix~\ref{app_mixing}, we describe the numerical implementation {of calculating the chemistry accounting for} 
the mixing rates.

The boundary conditions are set at the disk midplane and surface. Due to the symmetry with respect to the {midplane}, the net exchange rate
at $z=0$ is zero. Likewise, we assume that there is no exchange of material between the uppermost disk layer and any potential atmosphere on top of it.

\subsubsection{Vertical transport by disk winds}
We consider vertical disk winds as an alternative description of the vertical motion 
in the chemical model of the disk. Contrary to turbulent mixing, which
implies upward and downward motion, the disk wind transports material upwards 
only (c.\,f. Figure~\ref{fig_accretion_mixing_wind}(c) in Appendix~\ref{app_mixing}).

We refer to the wind solutions of \citet{suzuki09} to describe the vertical disk wind in our protoplanetary disk model.
They investigated disk winds driven by MHD turbulence in local three-dimensional MHD 
simulations of stratified accretion disks
and found that large-scale channel flows develop most effectively at $1.5$--$2$ disk 
scale heights, where the magnetic pressure is comparable to the gas pressure
(here, we refer to their model with initial plasma $\beta=10^{6}$).
The breakup of these channel flows drives structured disk winds, which might play an 
essential role in the dynamical dispersal of
protoplanetary disks. While \citet{suzuki09} perform local three-dimensional simulations at radius
$s=1\,\mathrm{AU}$ which lead to quasi-periodic cycles of $5$--$10$ rotations and strong mass outflows,
our disk model is steady and does not allow for a dispersal of the disk.

We therefore adopt their solutions for the wind velocity $\hat{v}_{\mathrm{wind},0}$ 
at the midplane at $1\,\mathrm{AU}$ and calculate the radial
and vertical profile according to \citet{suzuki09} and the vertical continuity 
equation in steady state, $\partial(\rho v_z)/\partial z = 0$,
\begin{eqnarray}
v_0&=& v_\mathrm{wind}(s,z\!=\!0) = \varkappa \hat{v}_{\mathrm{wind},0} 
\cdot \frac{c_\mathrm{s}(s)}{c_\mathrm{s}(1\,\mathrm{AU})},\nonumber\\
v_z&=&v_\mathrm{wind}(s,z) = v_0 \cdot \frac{\rho(s,z=0)}{\rho(s,z)}\,,\label{eqn_wind_velocity}
\end{eqnarray}
where $\hat{v}_{\mathrm{wind},0} = 8.5 \cdot 10^{-5} c_\mathrm{s}(1\,\mathrm{AU})$ \citep{suzuki09}.
For large values of $\hat{v}_{\mathrm{wind},0}$, the {midplane} layers 
disperse over timescales shorter than the typical disk lifetime of $10^6\mathrm{yr}$, which contradicts the assumption of a steady disk model.
We therefore introduce an ad-hoc factor $\varkappa=10^{-3}$ in Equation~\eqref{eqn_wind_velocity} 
to reduce the wind velocities such that the dispersal time scales,
$\Sigma/(\rho_\mathrm{gas}v_z)$ (where $\Sigma$ is the surface density of the disk),
become $\sim 10^{6}\mathrm{yr}$. The resulting wind profiles reproduce the
results of \citet{suzuki09} in the sense that the disk wind is most effective at 
$\sim 2$ disk scale heights, where the gas density and
therefore the gas pressure decreases steeply and any potential magnetic pressure would become comparable {with} it.

The wind rates are calculated from the vertical continuity equation for species $k$, 
using the wind velocity profile (Equation~\eqref{eqn_wind_velocity}):
\begin{equation}
\dot{n}_{k,\,\mathrm{wind}} = - \frac{\partial (n_k v_z)}{\partial z}\,.\label{eqn_wind_rate}
\end{equation}
The wind rates for the discrete grid of the disk are given in Appendix~\ref{app_wind}.

The boundary conditions are set as follows: since we adopt small wind velocities so
that the disk dispersal time becomes long enough, we can assume that
the mass loss is negligible at the {midplane} and therefore 
$\dot{n}_{k,\,\mathrm{wind}}=0$, whereas at the upper boundary,
we simply allow all material transported by the disk wind to be lost. 

\subsubsection{Timescales}\label{sec_timescales}

\begin{figure*}
\myincludegraphics{width=0.99\textwidth}{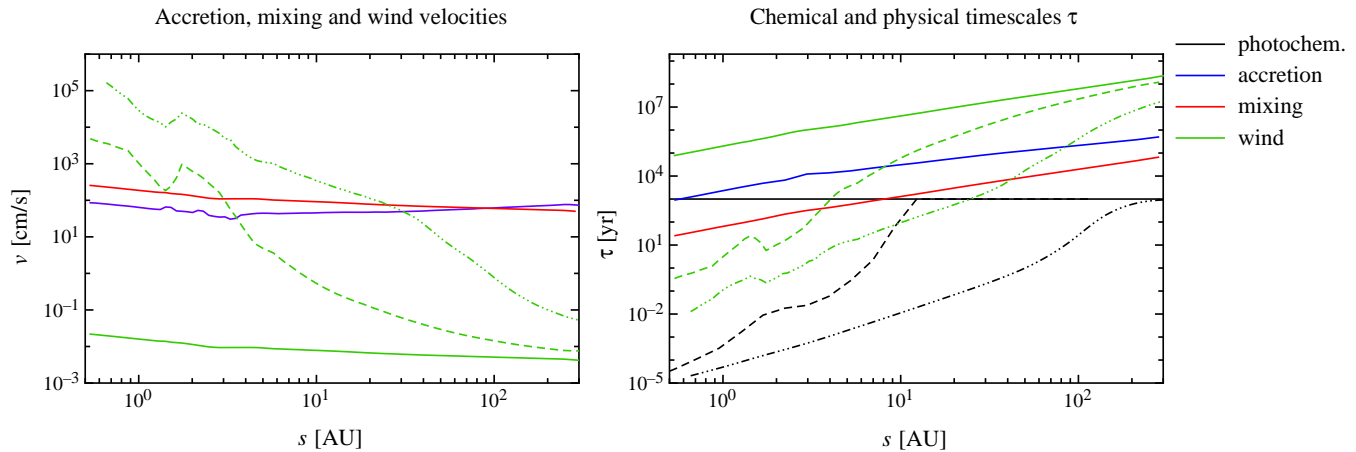}
\caption{Velocities (left) and timescales (right) for the transport processes of accretion (radial), mixing (vertical) and wind (vertical).
Typical timescales of the photochemical reactions are also shown in the right panel (see text for details). Solid lines denote midplane
values, i.\,e., $z/s=0$, (accretion, mixing, wind), or the timescale of the photochemistry for the interstellar radiation field (photochem.), respectively.
Dashed lines correspond to $z/s=0.2$ and dotted-dashed lines to $z/s=0.5$.}\label{fig_timescales_velocities}
\end{figure*}

The grid crossing times for the mass transport by viscous accretion, 
turbulent mixing and disk winds are displayed in 
Figure~\ref{fig_timescales_velocities}.
These are set by the CFL-condition for the discrete grid of our disk model:
\[
\tau_\mathrm{acc} = \frac{\Delta s}{v_s},\quad 
\tau_\mathrm{mix} = \frac{\Delta z}{v_{z,\,\mathrm{mix}}},\quad 
\tau_\mathrm{wind} = \frac{\Delta z}{v_{z,\,\mathrm{wind}}}.
\]
The radial and vertical size of the grid cells are given by $\Delta s$
and $\Delta z$, respectively, and
depend on distance $s$ from the star. For reference, we display
the accretion, mixing and wind velocities in Figure~\ref{fig_timescales_velocities} (left panel).

In the right panel of Figure~\ref{fig_timescales_velocities}, we also plot the timescale of photochemistry
given by \citep{duley84}:
\begin{equation}
\tau_\mathrm{photochem} = 10^3\mathrm{yr} \cdot \frac{F_\mathrm{FUV,ISRF}}{F_\mathrm{FUV}(s,z)}\,,
\end{equation}
where $F_\mathrm{FUV,ISRF}$ is the FUV photon flux of the interstellar radiation field.
In the upper layers of the disk, the UV radiation field is strong and
the timescale for the photochemistry is shorter than the dynamical timescales. 
Below the transition layer, $z/s\lessapprox 0.2$, 
the incident FUV radiation is highly attenuated and the chemical timescales are controlled by two-body 
reactions, which are sensitive to the temperature and density. The chemical timescale in this region
is generally longer than the photochemical timescale for the interstellar radiation field, $10^3\mathrm{yr}$. 
The accretion and mixing timescales become comparable to this limit at $0.5$ and 
$10\,\mathrm{AU}$, respectively. The disk wind timescales are longer than the chemical timescales for all
$z/s$ for the entire disk in this model.

\subsection{Infrared molecular line spectra}\label{sec_line_spectra}
Using the results of our chemical evolution calculations, we compute emergent mid-infrared spectra from the disk. 
We include the dust continuum, which is the dominant source of opacity in protoplanetary disks, along with
molecular line emission and absorption, \reftext{assuming local thermodynamic equilibrium (LTE)}.

The dust continuum opacities are taken from the Jena opacity tables, 
using routines contained within the RATRAN package \citep{ossenkopf94,hogerheijde00}.
In accordance with the dust properties of the disk model, we use the tables for 
spherical dust grains (parameter ``e5'' in the RATRAN routines) with thin ice coating (parameter ``thin'').

The HITRAN'2004 molecular spectroscopic database \citep{rothman05} provides a 
large set of energy levels and line transitions in the infrared
for 39 molecules, focussing on terrestrial atmospheric species. 
Nevertheless, the overlap of species of interest between the HITRAN
and UMIST databases is sufficient for our purposes. Since isotopes
are not currently included in our chemical reaction network, 
we neglect spectroscopic information for minor isotopes in the HITRAN database.

The major issue concerning the calculation of line intensities in the infrared is the limited 
availability of collisional coefficients, in particular,
for the large set of levels and transitions contained in the HITRAN database 
(e.\,g., more than $180000$ transitions for the main isotope of
CH$_4$ for wavelengths between $1$ and $40\,\micron$). 
Since the main purpose of this
work is the calculation of the disk's chemical evolution, a detailed non-LTE radiative 
transfer calculation of the emergent spectra is beyond
the scope of this paper. We therefore assume local thermal equilibrium to model the line emission and 
absorption strengths for the species of interest. 
For a species
\emph{k}, the transition $j\to i$ ($E_i<E_j$) is calculated from the standard equations
\begin{eqnarray}
Q_k(T_\mathrm{gas}) &=& \sum_i g_{k,i} \exp\left(-\frac{E_{k,i}}{T_\mathrm{gas}}\right)\,,\\
n_{k,i/j}      &=& n_k \frac{g_{k,i/j}}{Q_k(T_\mathrm{gas})} \cdot \exp\left(-\frac{E_{k,i/j}}{T_\mathrm{gas}}\right)\,,\\
j_{k,ij}       &=& h \nu A_{k,ji} n_{k,j} \mathcal{V}_k\,,\phantom{\int_A^B}\\
\kappa_{k,ij}  &=& h \nu (B_{k,ij} n_{k,i} - B_{k,ji} n_{k,j}) \mathcal{V}_k\phantom{\int_A^B}.
\end{eqnarray}
Here, $n$, $j$ and $\kappa$ denote the level populations, the line emissivity and the line opacity, respectively.
The Einstein coefficients $A$ and $B$ are listed in the HITRAN database 
or determined using the principle of detailed balance. The database also
provides statistical weights $g$ and partition functions $Q$ for a wide range of temperatures. 
We calculate Voigt profiles $\mathcal{V}$
including natural line broadening, pressure broadening and thermal broadening.
A Doppler profile at a temperature, $T_\mathrm{gas}$, is assumed for the thermal broadening.
\reftext{Turbulent broadening is neglected, since it is negligible compared to thermal broadening for subsonic turbulent velocities.}
The half-widths for natural broadening $\gamma_\mathrm{n}$
and pressure broadening $\gamma_\mathrm{p}$ are provided by the 
HITRAN database at a temperature of $296\,\mathrm{K}$.
The temperature dependence of pressure broadening is approximated by
\begin{equation}
\gamma_\mathrm{p} = \gamma_\mathrm{p}(296\,\mathrm{K}) \cdot \left(\frac{296\,\mathrm{K}}{T_\mathrm{gas}}\right)^{n_\mathrm{p}}\!\!,
\end{equation}
where $n_\mathrm{p}$ is the temperature-dependent exponent in the HITRAN database.

We adopt a simple, one-dimensional line-of-sight integration to calculate the emergent disk spectra for
a disk viewed face-on, i.\,e., with an inclination $i=0$, relative to the observer.
The radial resolution for the spectral calculation is identical to the chemical model. 
Along each ray, the intensity
is integrated following the one-dimensional equation of radiative transfer, accounting for the thermal dust continuum
emission and the molecular line emission,
\reftext{\begin{equation}
\frac{dI_{\nu'}}{dz}=(\nu'/\nu)^2 \Bigl[j_{\nu} - \kappa_{\nu} I_{\nu}\Bigr]\,.\label{eqn_radiative_transfer}
\end{equation}}%
\reftext{Quantities on the right hand side of \eqref{eqn_radiative_transfer} are evaluated 
in the moving frame of a fluid element of the observed object, while those
on the left hand side are given in the rest frame of the observer (denoted with $'$).
The observed frequency is given by
$\nu'=\nu(1+(v/c)\cos\theta)$, where $v$ is the velocity of the fluid
element and $\theta$ is the angle between the directions of the velocity
and the observer.} The emission and absorption coefficients are given by
\begin{eqnarray}
j_\nu&=&\sum_k \sum_{E_i<E_j} j_{k,ij} + \kappa_{\nu,\mathrm{dust}}B_\nu(T_\mathrm{dust})\,,\\
\kappa_\nu&=&\sum_k \sum_{E_i<E_j} \kappa_{k,ij} + \kappa_{\nu,\mathrm{dust}}\,,
\end{eqnarray}
where $B_{\nu}$ is the \reftext{Planck} function and $\kappa_{\nu,\mathrm{dust}}$ is the specific dust absorption.

The total emission from the disk is calculated as
\reftext{\begin{equation}
L_{\nu',\,\mathrm{disk}} = 8\pi^2\int s\,ds\,I_{\nu'}(s)\,.
\end{equation}
The full spectrum is obtained from
\begin{eqnarray}
L_{\nu'}&=&L_{\nu',\,\mathrm{star}} + L_{\nu',\,\mathrm{disk}}\phantom{\int}\nonumber\\
&=&4\pi^2 R^2 \cdot B_{\nu'}(T_\mathrm{eff}) + L_{\nu',\,\mathrm{disk}}\,.\label{eqn_full_spectrum} 
\end{eqnarray}}
Here, the stellar emission is modeled as blackbody emission with temperature $T_\mathrm{eff}=4000\,\mathrm{K}$
and stellar radius $R = 2 R_\odot$ and is added to the dust continuum and molecular line emission from the disk.

\reftext{For the line-of-sight integration, a volume-weighting method is adopted to interpolate the temperature
and molecular densities before computing the emission and absorption coefficients. In doing so, potential effects of the
relatively coarse grid on the emerging line emission are reduced.}
\section{Results and discussion}\label{sec_results}
\begin{figure*}
\myincludegraphics{width=0.94\textwidth}{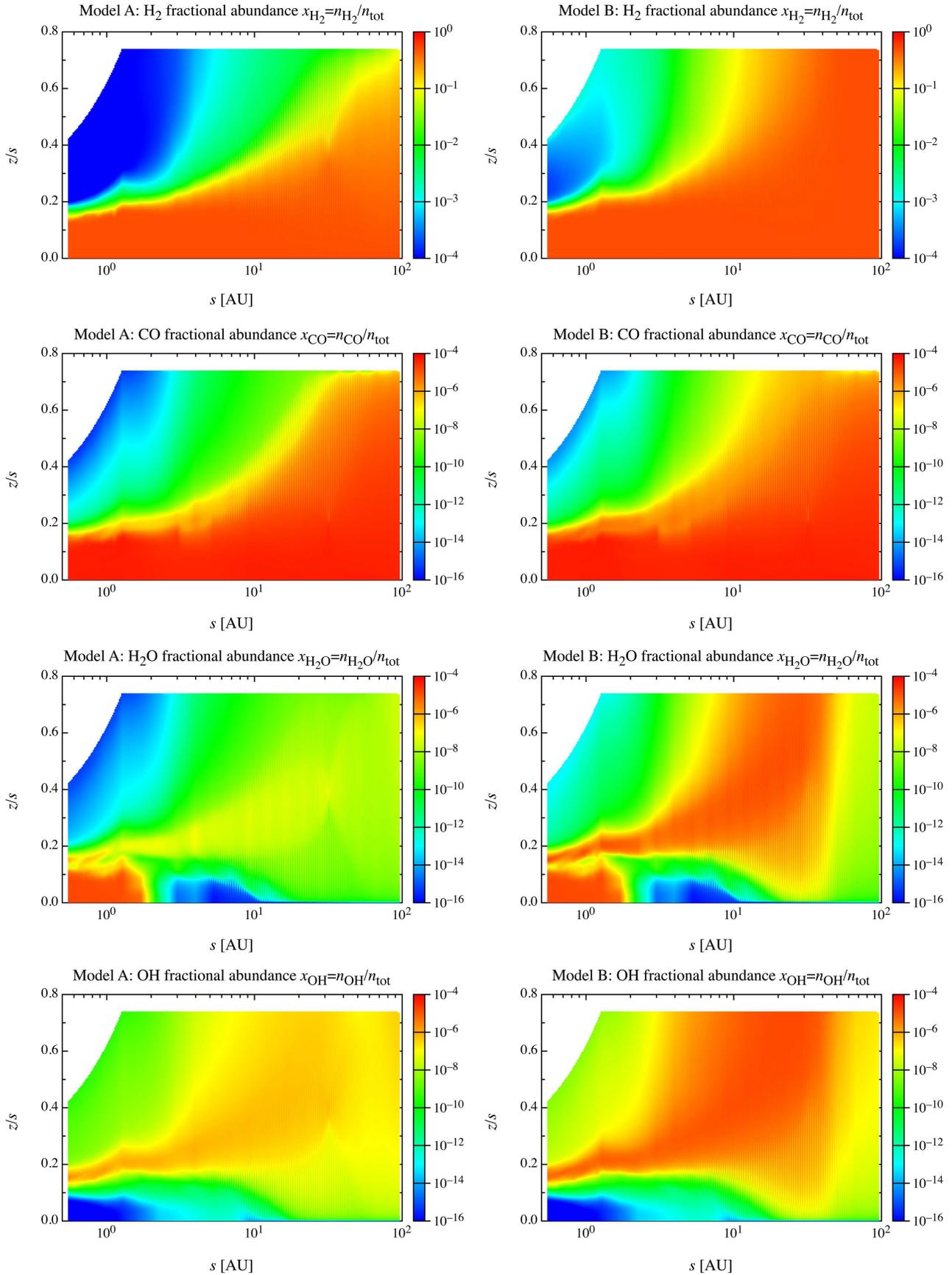}%
\caption{Gas-phase fractional abundances of H$_2$, CO, H$_2$O and OH for models A and B.}\label{fig_grain_formation_1}
\end{figure*}

In the following, we discuss the disk chemical evolution results and spectral calculations 
for the different models considered in this study. 
First, we investigate the influence of the H$_2$ formation rates on grains on the disk
chemistry (Section~\ref{sec_grain_formation}). In Sections~\ref{sec_results_interaction_1} and 
\ref{sec_results_interaction_2},
we study the specific effects of viscous accretion, turbulent mixing and disk winds.
Finally, in Section~\ref{sec_discussion_observations} we discuss how our results 
compare with observations of protoplanetary disks.

\subsection{H$_2$ grain formation rates}\label{sec_grain_formation}

Recently, \citet{glassgold09} studied the effect of the H$_2$ formation rates on dust grains
on the water abundances at the disk surface of the inner disk. We revisit this effect and also
study the influence on the infrared molecular line spectra.

In model A, we use the simple approximation from \citet{gould63} and \citet{hollenbach71}. 
In model B, the adopted H$_2$ grain formation rate is that
according to \citet{cazaux02} (see Section~\ref{sec_gas_grain_interaction}).
In both models, the effects of accretion, turbulent mixing and disk winds are ignored. 
The chemical evolution of the disk is calculated
over $10^6\mathrm{yr}$ for a radial range of $0.5$--$300\,\mathrm{AU}$. 
For both models, the chemical abundances everywhere in the disk have reached steady state by approximately 
$10^5\mathrm{yr}$.

In Figure~\ref{fig_grain_formation_1}, we show the resulting gas-phase abundances
for molecular hydrogen, carbon monoxide, water and the hydroxyl radical
for the inner $100\,\mathrm{AU}$ of the disk.
In model B, the formation of molecular hydrogen is more efficient, particularly in the 
upper layers of the disk. In the warm molecular layer,
i.\,e., the transition layer where the temperature rises and the density decreases steeply, 
H$_2$ is prone to continuous dissociation due to
thermal reactions and incident radiation, and the enhanced grain formation rates in model B 
are able to maintain higher fractional abundances.
Similarly, in the \reftext{photodestruction} region, i.\,e., the surface layers of the disk,
H$_2$ is more abundant by a factor of $10$--$50$ between $5$ and $100\,\mathrm{AU}$. 
At even smaller radii the strong stellar UV radiation destroys H$_2$ on timescales
shorter than it is formed even in model B. In the lower layers and outer regions of the disk, 
the differences between model A and B become negligible, since here, H$_2$ accounts for essentially all of the hydrogen nuclei.

The differences in the hydrogen chemistry also affect the resulting abundances of predominant 
molecules in the inner disk, particularly CO and H$_2$O.
These molecules are commonly found in the planet-forming regions of protoplanetary disks \citep[c.\,f.,][]{salyk08}. 
Their modeled abundances depend strongly on the amount of molecular hydrogen at the 
disk surface and are therefore 
very different for model A and B. The increase of molecular hydrogen
by a factor of $10$--$50$ between $5$ and $100\,\mathrm{AU}$ in the upper layers translates to
an increase in CO and H$_2$O by a factor of $\approx 1000$. 
Such a strong correlation is expected theoretically \citep{glassgold09}: high fractional abundances of H$_2$ of $\sim 10^{-3}$
in the surface layers produce significant amounts of CO, H$_2$O and OH through a sequence of neutral radical reactions:
\begin{eqnarray*}
\mathrm{O} + \mathrm{H}_2 &\ \longrightarrow\ & \mathrm{H} + \mathrm{OH}\,\\
\mathrm{OH} + \mathrm{H}_2 &\ \longrightarrow\ & \mathrm{H}_2\mathrm{O} + \mathrm{H}\,\\
\mathrm{C} + \mathrm{OH} &\ \longrightarrow\ & \mathrm{CO} + \mathrm{H}\,.
\end{eqnarray*}
\citet{glassgold09} show that the ratio $x(\mathrm{H}_2\mathrm{O})/x(\mathrm{O})$ depends quadratically on the ratio
$x(\mathrm{H}_2)/x(\mathrm{H}^+)$. Our results confirm this correlation, since the $\mathrm{O}$ and $\mathrm{H}^+$ 
abundances are practically identical between model A and B. 

\begin{figure*}
\myincludegraphics{width=0.96\textwidth}{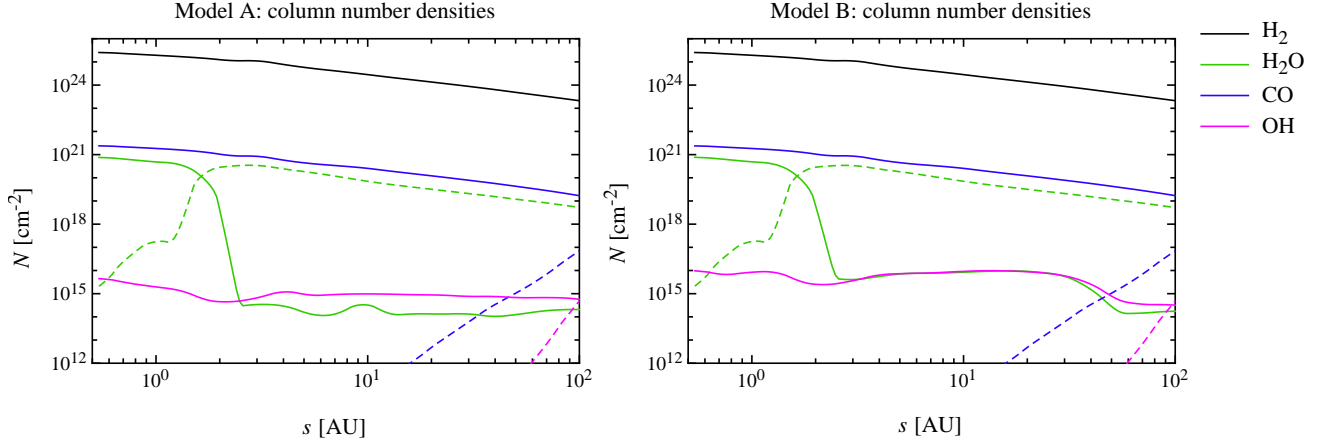}%
\caption{Column densities for the species of interest in models A and B. Solid lines correspond to gas-phase species, 
dashed lines to ices.}\label{fig_grain_formation_3}
\end{figure*}

The molecular column densities are only slightly changed compared with the fractional abundances, since the
the latter are altered only at the disk surface where the density is low (Figure~\ref{fig_grain_formation_3}). 
Independent of the H$_2$ formation rate, gaseous CO traces the total disk mass,
with a constant ratio $N_\mathrm{CO}/N_{\mathrm{H}_2}=10^{-4}$ at radii $<100\,\mathrm{AU}$ where the dust temperature
is high enough so that CO does not freeze out. Gas-phase water is fairly abundant at radii $<3\,\mathrm{AU}$ 
(i.\,e., within the snow line of the disk) with $N_{\mathrm{H}_2\mathrm{O}}/N_\mathrm{CO}=0.25$. 
The column density of water is increased by about one order of magnitude
between $5$ and $50\,\mathrm{AU}$, which is due to the large overabundance in the upper layers of model B.

\begin{figure*}
\myincludegraphics{width=0.99\textwidth}{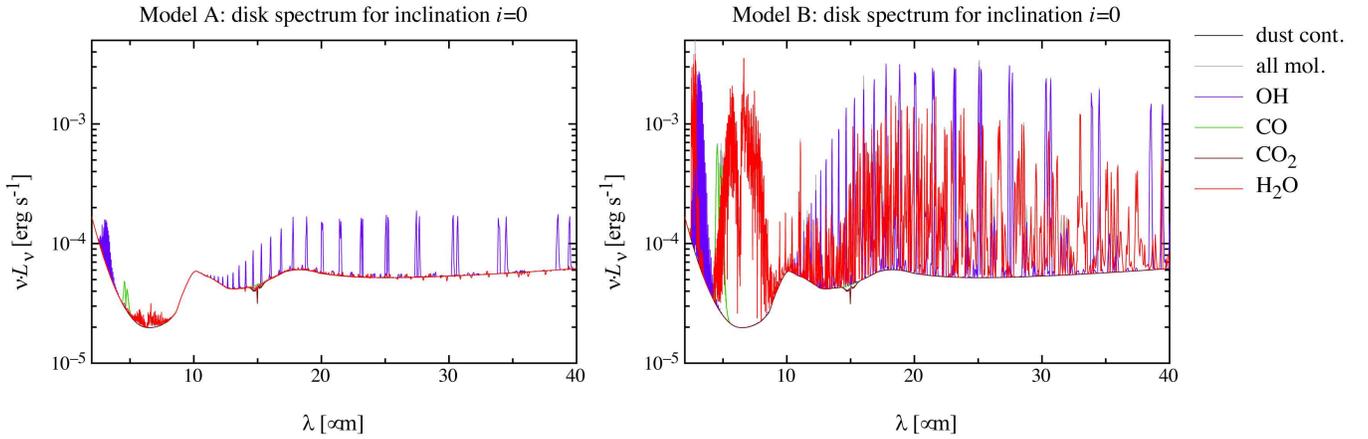}\hfill
\caption{Emergent \reftext{LTE} disk spectra for models A and B for dust continuum and molecular line emission.}
\label{fig_grain_formation_4}
\end{figure*}

The two H$_2$ formation rate models affect the abundances of the observable molecules 
in the warm molecular layer and the \reftext{photodestruction} layer,
from where infrared molecular line emission originates. 
In Figure~\ref{fig_grain_formation_4}, 
we display the emergent spectra
for a disk seen face on (inclination $i=0$). The spectral resolution is set to $600$ 
to match the Spitzer/IRS specifications. The molecular line emission is calculated
for the following molecules between $1$ and $40\micron$: 
CO, H$_2$O, OH, CO$_2$, CH$_4$, NO, SO$_2$, NH$_3$, C$_2$H$_2$, N$_2$, H$_2$S, and NO$^+$.
The spectra differ substantially between the two models. Model A produces 
strong and narrow emission lines of OH between $10$ and $40\micron$,
due to high abundances in the surface layers of the disk. On the contrary, 
H$_2$O shows broad emission bands between $5$ and $10\micron$, but only
weak and narrow emission and absorption features at larger wavelengths, in 
contrast to the observations \citep[c.\,f., {Figure}~1 in][]{carr08}.
The CO emission band at $4$--$5\micron$ is present, as well as an absorption 
dip of CO$_2$ at $15\micron$. The absorption
signatures are generated at radii $\lessapprox 2\,\mathrm{AU}$, where the temperature 
increases near the midplane due to viscous heating
(c.\,f., the inlay in Figure~\ref{fig_disk_properties}). Since the abundances of CO$_2$ and H$_2$O 
in model A are high near the disk midplane, but low 
in the disk surface where the gas temperature increases towards an observer, absorption 
lines rather than emission lines are generated. 
Emission or absorption from other molecules included in the spectral calculation (e.\,g., NH$_3$, CH$_4$) 
is very weak in the spectrum. 

Model B, on the other hand, produces a spectrum with enormous emission features from OH, H$_2$O and CO.
The CO$_2$ absorption feature is not affected by the H$_2$
formation model, since it is destroyed very efficiently in the upper layers of the disk, 
where the main differences in the molecular abundances arise
between model A and B. Comparing the two-dimensional abundances of OH, H$_2$O and CO
in model A and B (Figure~\ref{fig_grain_formation_1}) allows us to constrain the origin of the emission lines to
roughly $3$--$20\,\mathrm{AU}$, where the gas temperature is high enough to contribute to the total flux
(c.\,f., Figure~\ref{fig_disk_properties}). 
Such strong emission features are certainly in disagreement with observations. 
The enhanced formation rate of H$_2$ on grains from \citet{cazaux02} might simply produce too much 
H$_2$ due to the choice of the variety of parameters used for this 
model, however, we should note here the limitations of our simple dust model. 
Firstly, we assume that 
gas and dust are well mixed in the disk, with a constant
mass density ratio of \reftext{$\rho_{gas}/\rho_{dust} \approx$ 100}. Numerous studies show that dust grains will settle towards the 
midplane instead of residing in the upper
disk layers \citep[see, e.\,g.,][for a review]{dominik07}.
Less dust grains in these layers lead to lower H$_2$ abundances and thus to lower abundances of 
water and OH. \citet{aikawa06}
demonstrated that less dust grains will also lead to a lower gas temperature at the disk surface, 
which reduces the line intensities.
Since the same line of argument applies to model A, it implies even lower abundances and thus 
weaker emission lines for H$_2$O in this model.
Secondly, the spectral calculations in
our study assume LTE level populations, which may overestimate the 
strength of the emission lines. \citet{meijerink09} studied the
effects of non-LTE radiative transfer on the molecular emission lines of water: they assume fractional H$_2$O abundances
between $10^{-10}$ and $10^{-4}$ throughout the disk. Interestingly, they
conclude that a strong depletion at $s<1\,\mathrm{AU}$ is required to account for observed emission line spectra, 
and such a depletion
is found in our model A. There, the H$_2$O abundance is low in the surface layer, because of the efficient 
destruction of water due to the stellar irradiation and the high
temperature. Most important, however, is that \citet{meijerink09} find drastic differences between 
LTE and non-LTE spectral calculations: the LTE calculations
overestimate the strength of the emission lines by a factor of $10$ for the strongest transitions 
(i.\,e., the largest Einstein coefficients). In model B, the
line emission is generated in the upper layers of the disk, where the gas number density is of the order 
of $10^{8}\mathrm{cm}^{-3}$.
\reftext{The water line transitions between $10$ and $40\micron$ reported from Spitzer
observations~\citep[e.\,g.,][]{carr08} are mainly rotational transitions in the ground vibrational state
and in the first vibrational state with critical densities of $10^{12}\mathrm{cm}^{-3}$ or larger \citep{meijerink09}.
Vibrational transitions between the two states that fall into this wavelength range have even higher critical densities.}
Hence, the conditions for LTE are partly violated and a proper non-LTE calculation of the disk spectra 
should be considered. In the following sections, we ignore the enhanced H$_2$ formation rates of model B and study the 
effects of accretion, turbulent mixing and wind on model A only.

\subsection{Accretion, mixing, disk winds -- single column solutions}\label{sec_results_interaction_1}
To investigate the effects of including physicochemical interaction, we first focus on one vertical column 
of the disk at radius $s=1.3\,\mathrm{AU}$.
Here, the temperatures of gas and dust are high enough, in particular in the upper layers of the disk, 
to allow for an active chemistry. 
Further, near- and mid-infrared observations derive characteristic radii for molecular line emission around $1\,\mathrm{AU}$ 
(c.\,f., Section~\ref{sec_discussion_observations}).

\begin{figure}
\myincludegraphics{width=0.99\columnwidth}{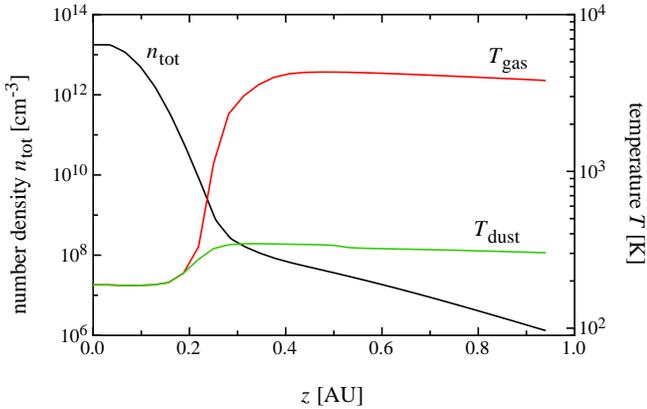}
\caption{Gas and dust temperature $T_\mathrm{gas}$ and $T_\mathrm{dust}$, and total number density 
$n_\mathrm{tot}$ as function of vertical distance
from the disk {midplane} at {a} radius $s=1.3\,\mathrm{AU}$.}\label{fig_T_and_n_at_is18}
\end{figure}

In the following, we discuss in detail the modeled molecular abundances for a number of species of interest. 
For a better understanding, in particular for
interpreting the resulting fractional abundances $x_k = n_k/n_\mathrm{tot}$, we display the 
physical properties of the disk column
at this radius in Figure~\ref{fig_T_and_n_at_is18}. 
The gas and dust temperatures are tightly coupled in the dense midplane with values around 
$200\,\mathrm{K}$. Hence,
practically all species have desorbed from the dust grains and the ice abundances are negligible. 
In the warm molecular layer ($z\approx 0.2\,\mathrm{AU}$), the gas and dust temperatures
begin to decouple due to the lower density. In the \reftext{photodestruction} layer, the gas is heated up to
\reftext{$4000\,\mathrm{K}$} because of the strong stellar irradiation, 
while the dust temperature is \reftext{around} a factor of $10$ lower.

In Figure~\ref{fig_interesting_species_at_is18_part_1}, we plot the fractional abundances as a function
of vertical distance from the midplane for the model without physicochemical interaction (model A) and for models with
either accretion, turbulent mixing or disk winds (models ACR, MIX, WND). 
The gas-phase abundances are shown together with the -- mostly negligible -- 
ice abundances.
\reftext{All four models make use of the simple H$_2$ formation rates on grains 
(Eq.~\eqref{eqn_h2_grain_formation_model_A}), while model B (not displayed)
includes the enhanced H$_2$ formation rates on grains 
(Eq.~\eqref{eqn_h2_grain_formation_model_B}) without physicochemical interaction.}
Table~\ref{tab_column_densities} at the end of this section compares the column densities of the most important species
at radius $1.3\,\mathrm{AU}$ for all models A, B, ACR, MIX, WND. In the following, we discuss in detail the effects of turbulent mixing
on the vertical chemical stratification at a radius $s=1.3\,\mathrm{AU}$, since this model
has the largest effects on the chemical structure of the disk.
\subsubsection{Turbulent mixing}
\begin{figure*}
\myincludegraphics{width=0.42\textwidth}{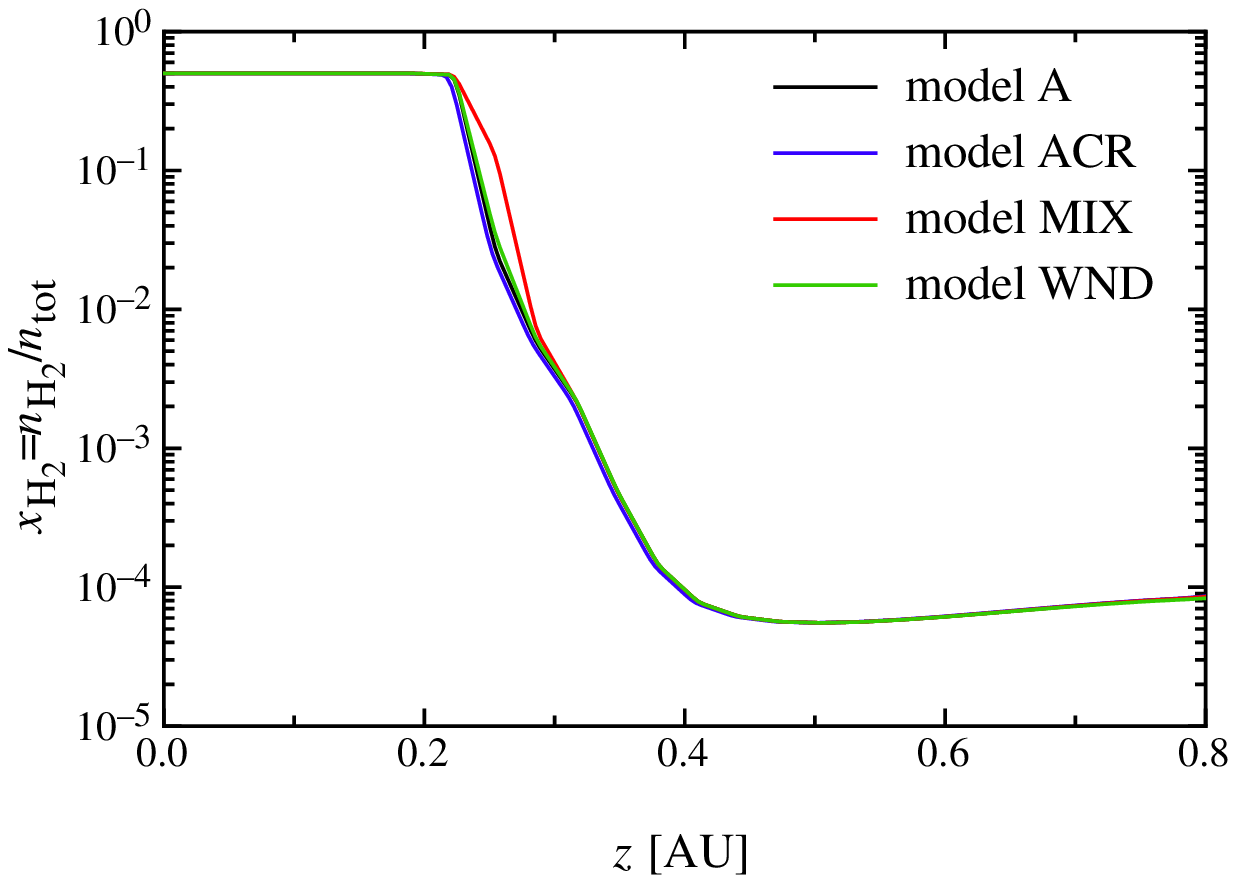}\hfill%
\myincludegraphics{width=0.42\textwidth}{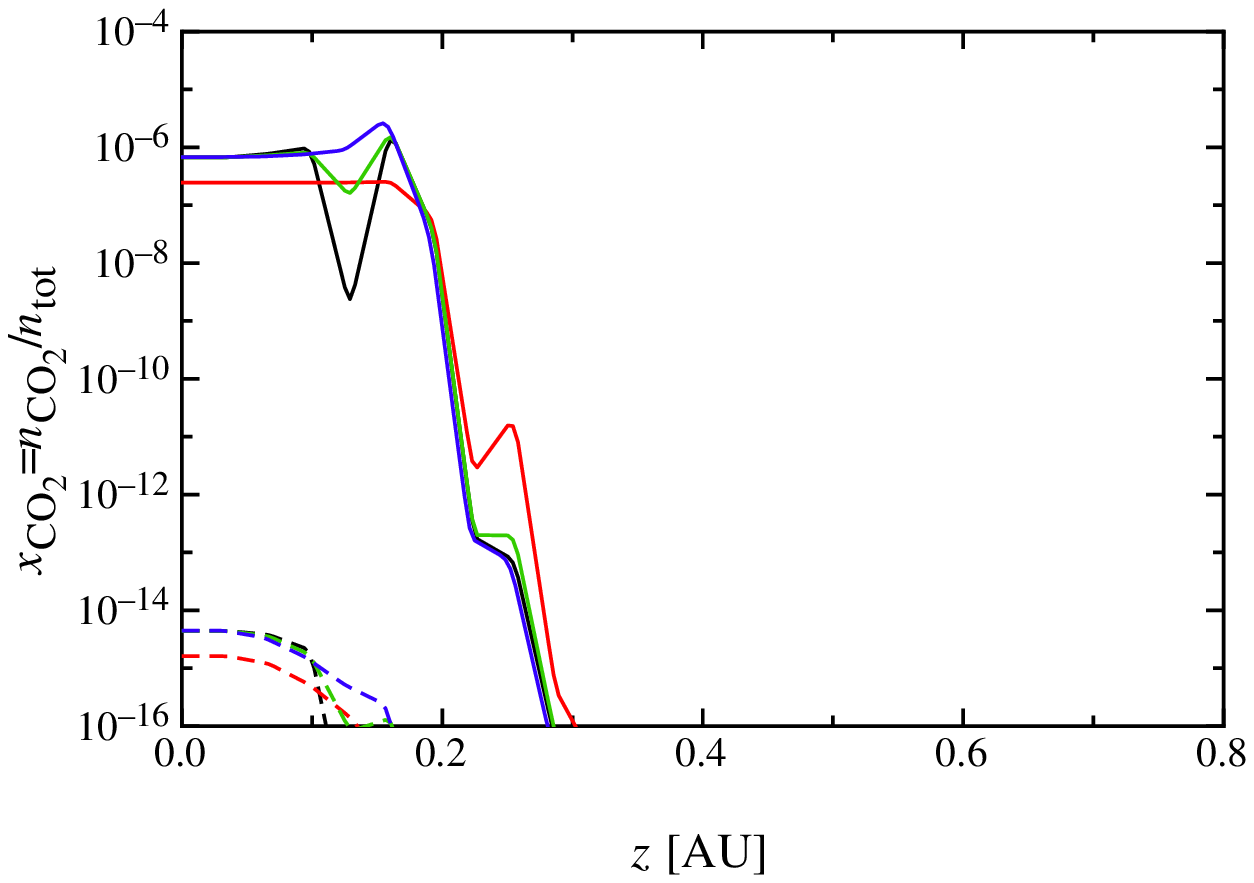}\\
\myincludegraphics{width=0.42\textwidth}{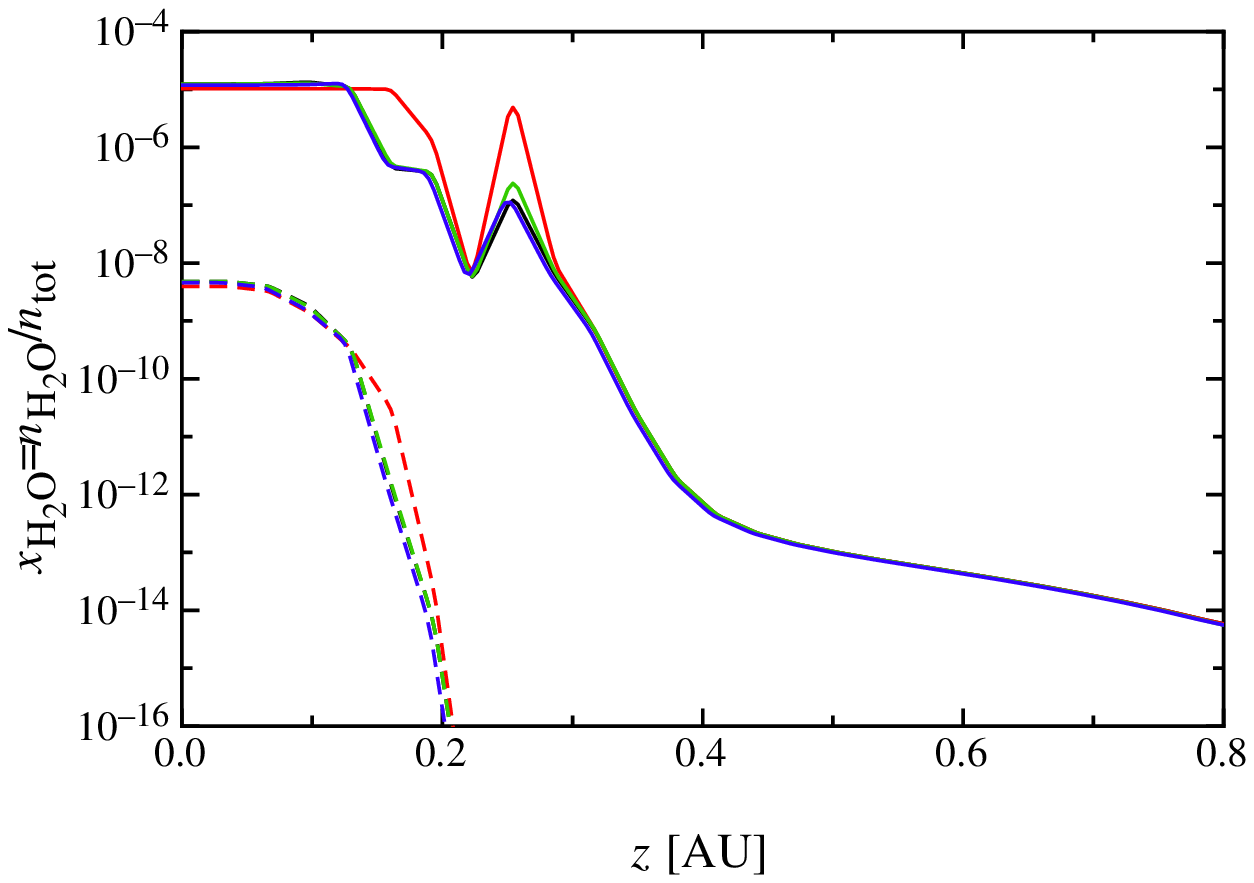}\hfill%
\myincludegraphics{width=0.42\textwidth}{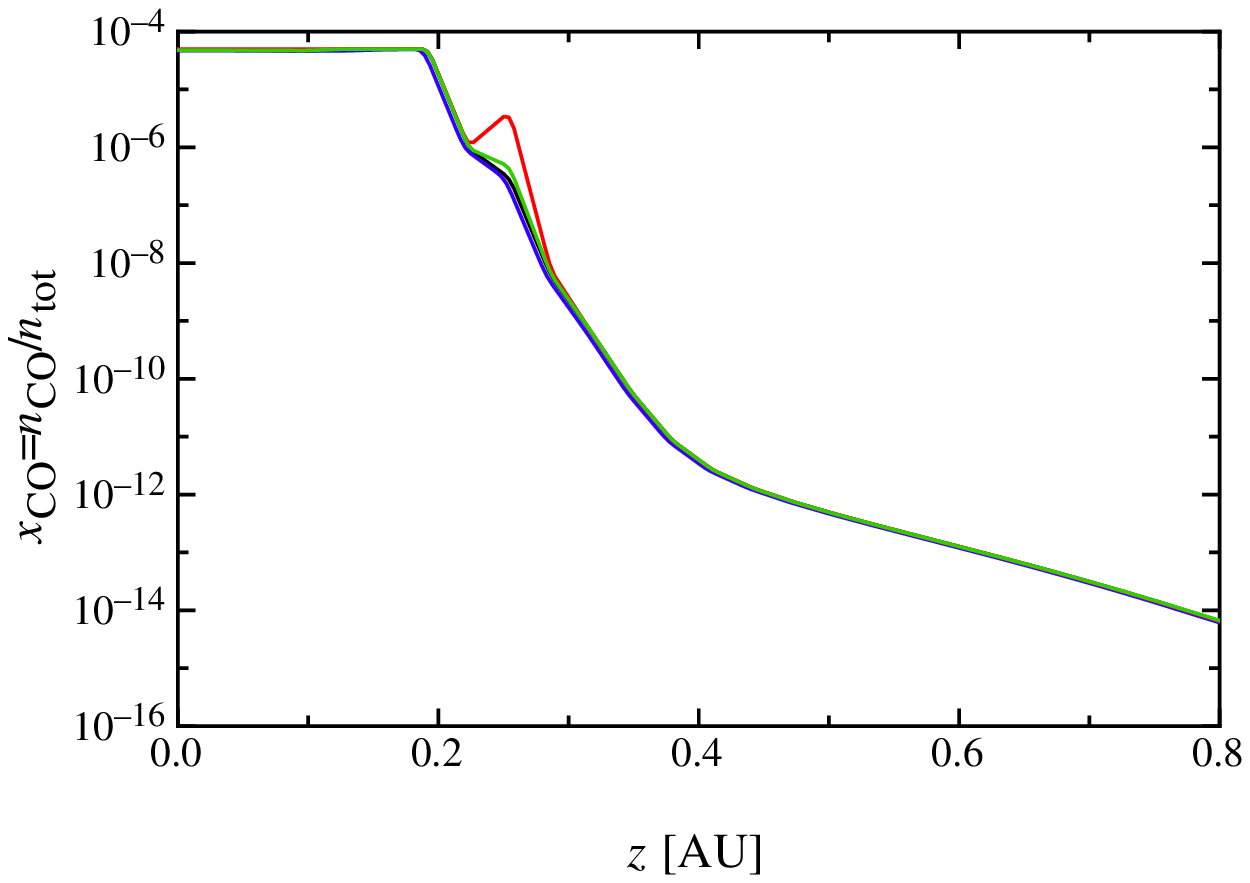}\\
\myincludegraphics{width=0.42\textwidth}{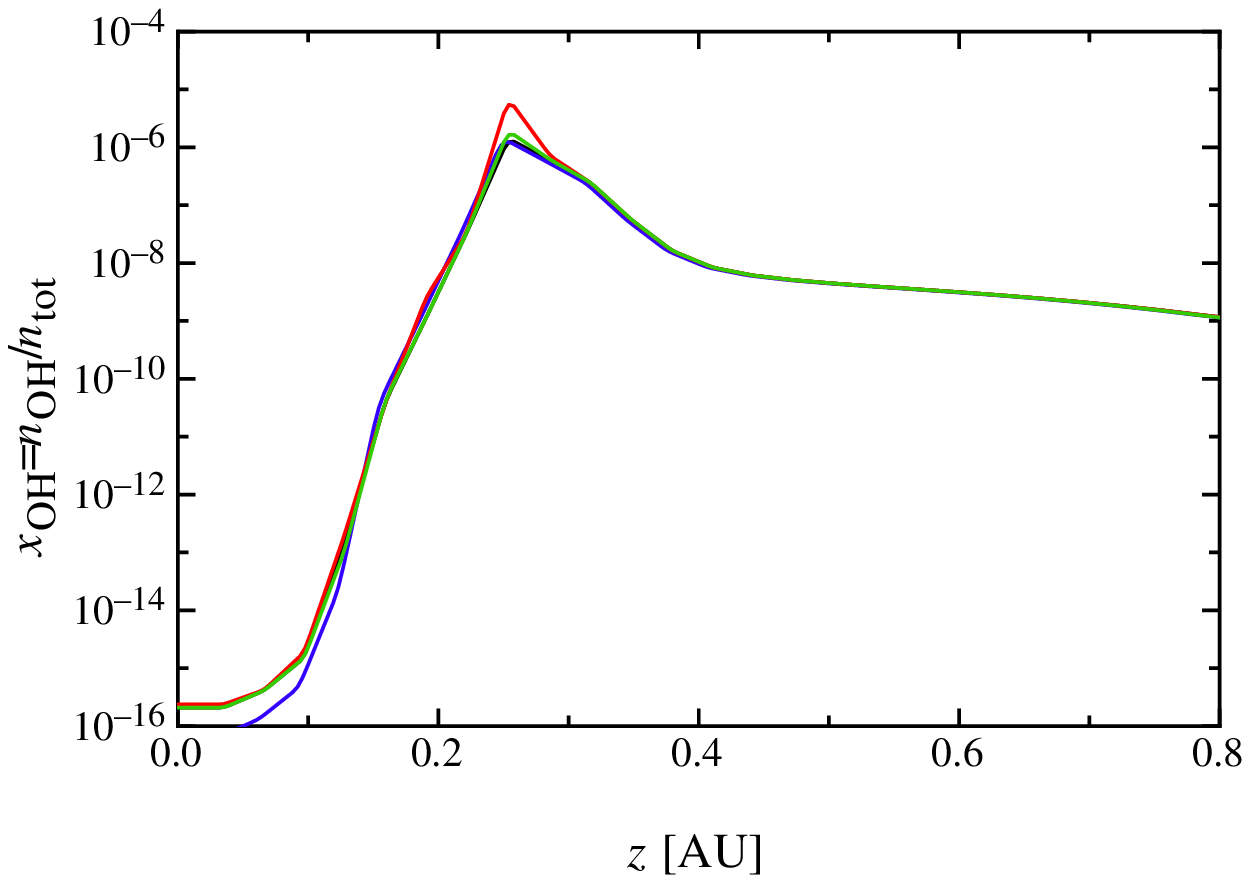}\hfill%
\myincludegraphics{width=0.42\textwidth}{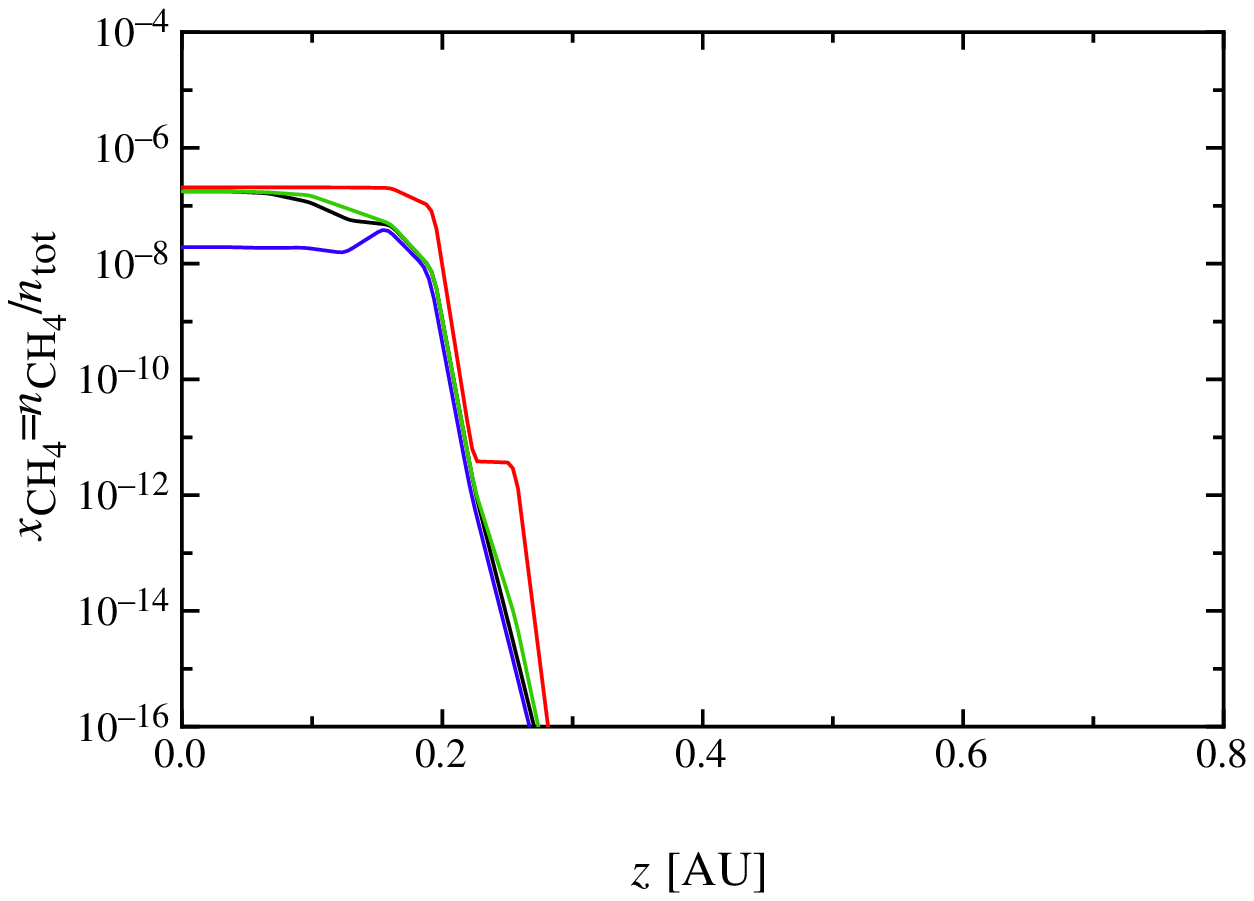}\\
\myincludegraphics{width=0.42\textwidth}{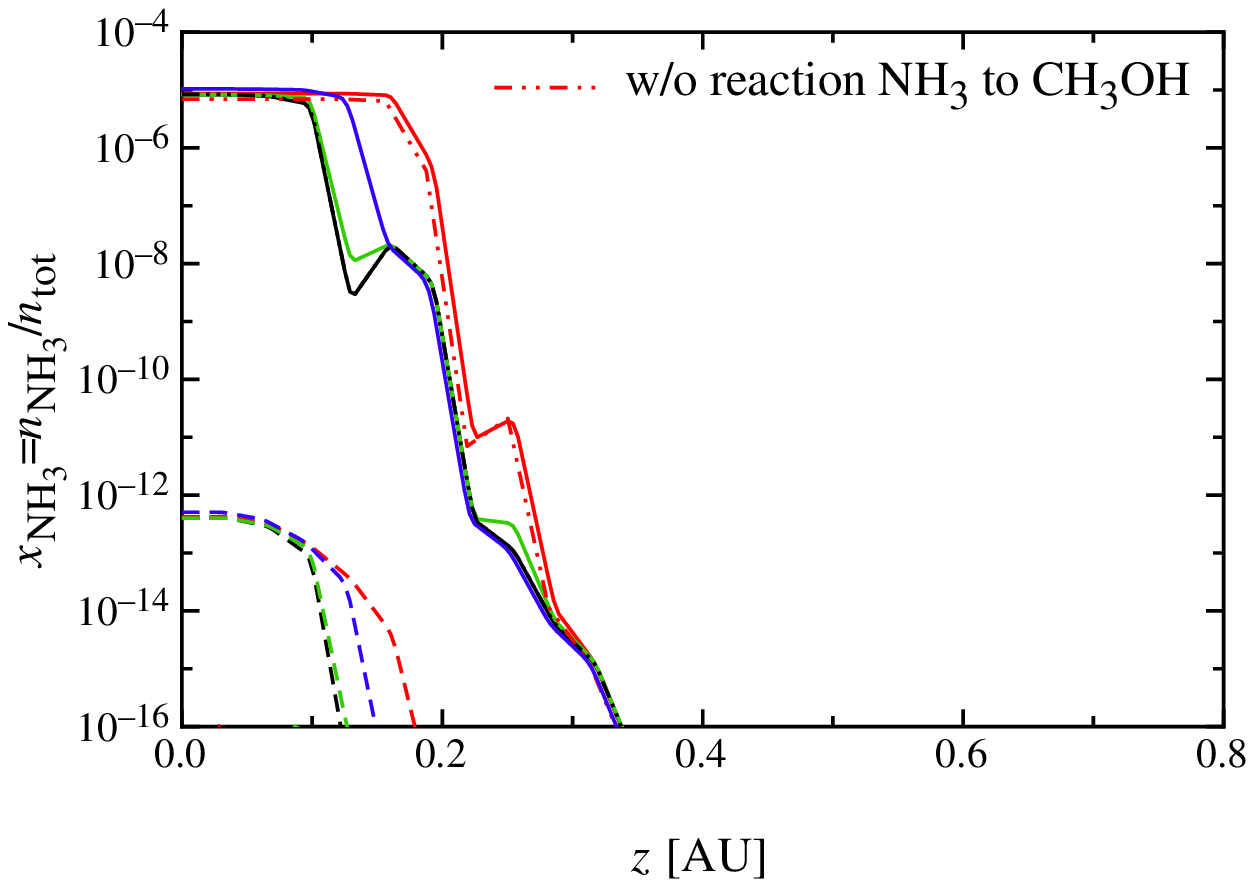}\hfill%
\myincludegraphics{width=0.42\textwidth}{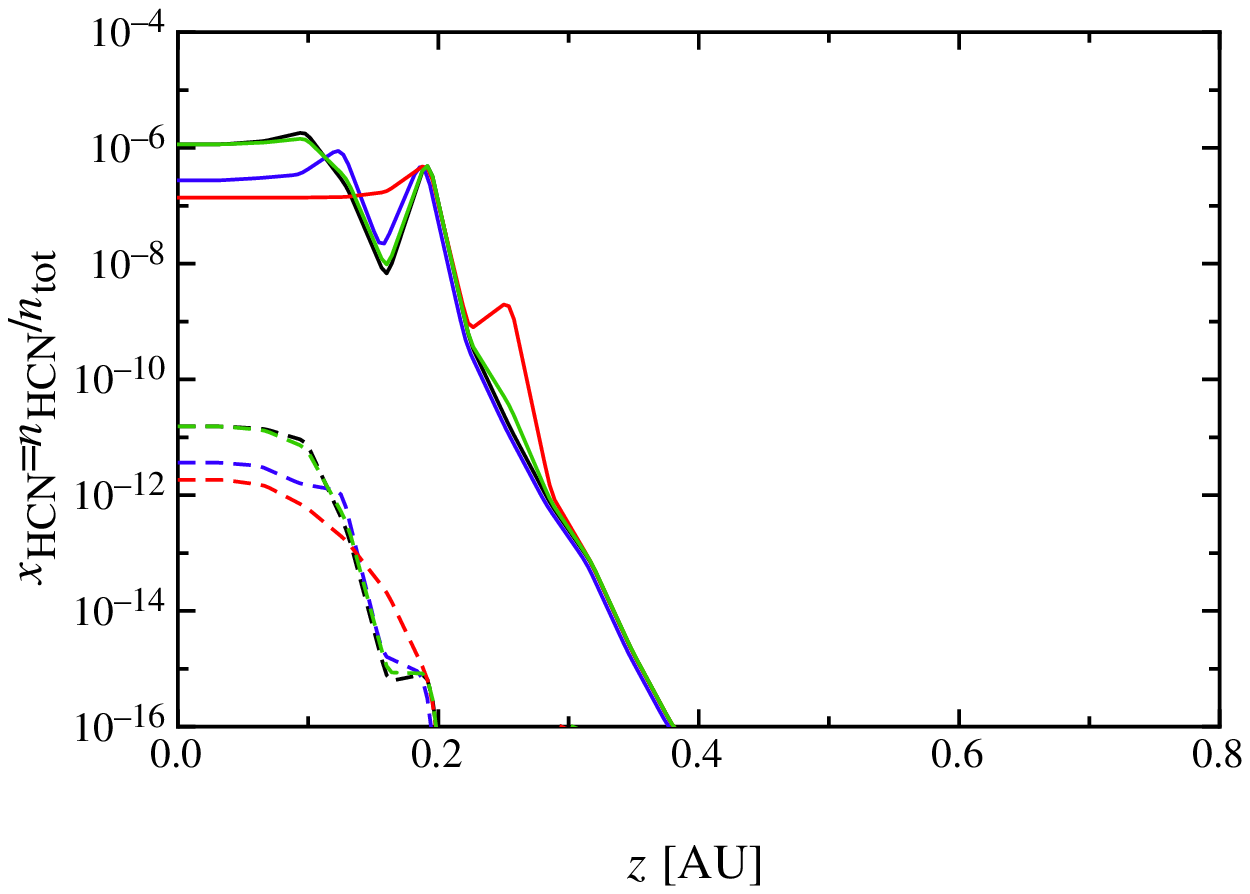}%
\caption{Fractional abundance profiles of gas-phase species (solid) and ices (dashed) 
as a function of vertical distance from the midplane at $s=1.3\,\mathrm{AU}$ in model A (black),
and in models with accretion (ACR, blue), mixing (MIX, red) or disk winds (WND, green).}
\label{fig_interesting_species_at_is18_part_1}
\end{figure*}
\addtocounter{figure}{-1}
\begin{figure}
\myincludegraphics{width=0.42\textwidth}{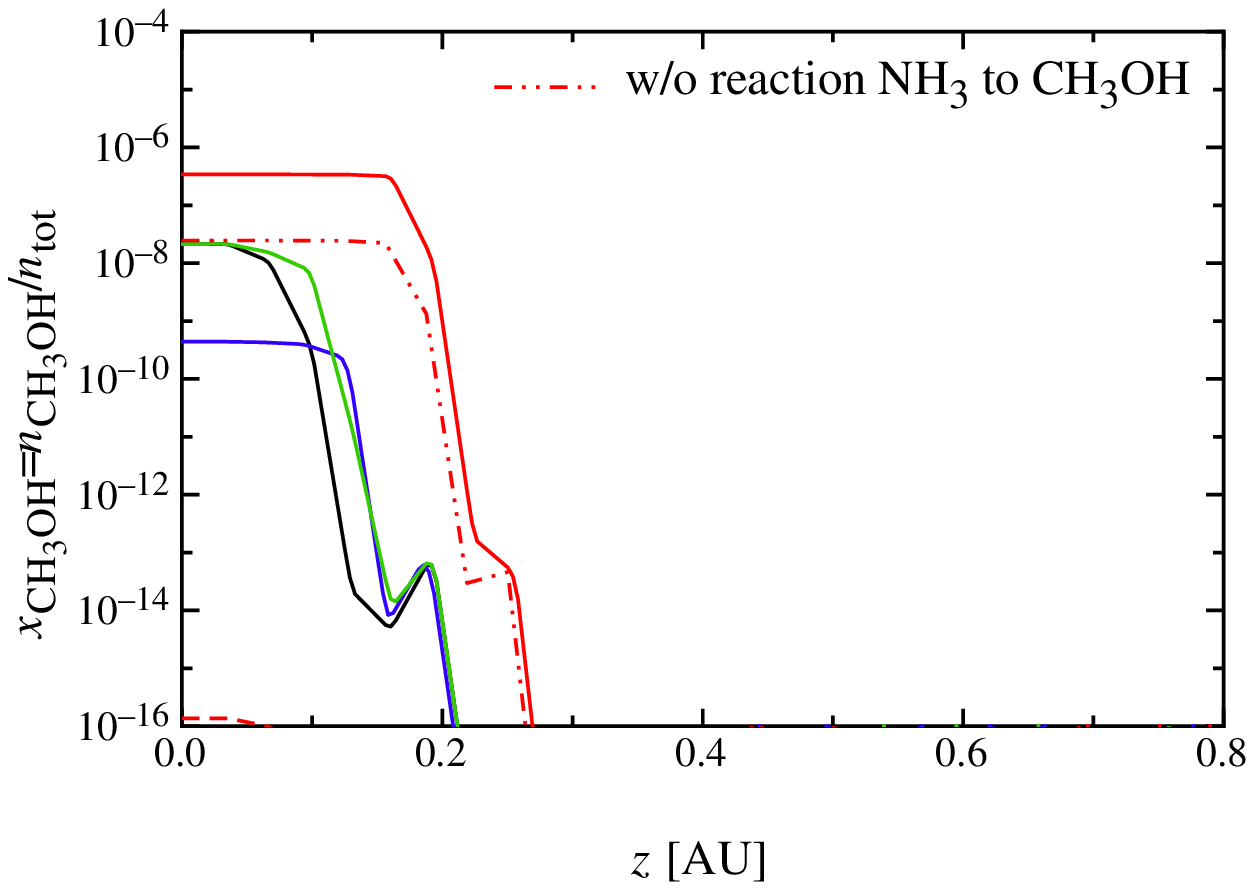}\\
\myincludegraphics{width=0.42\textwidth}{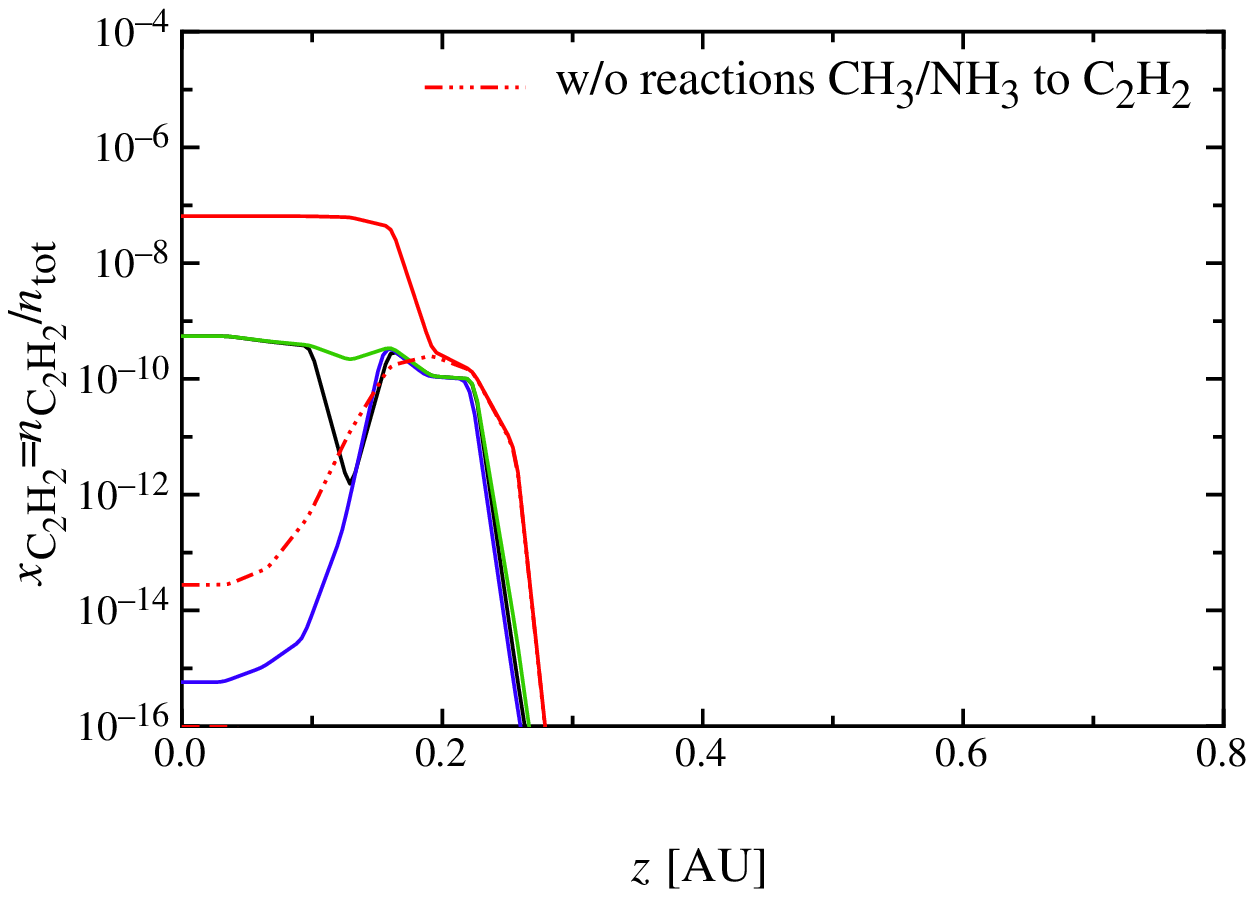}\\
\myincludegraphics{width=0.42\textwidth}{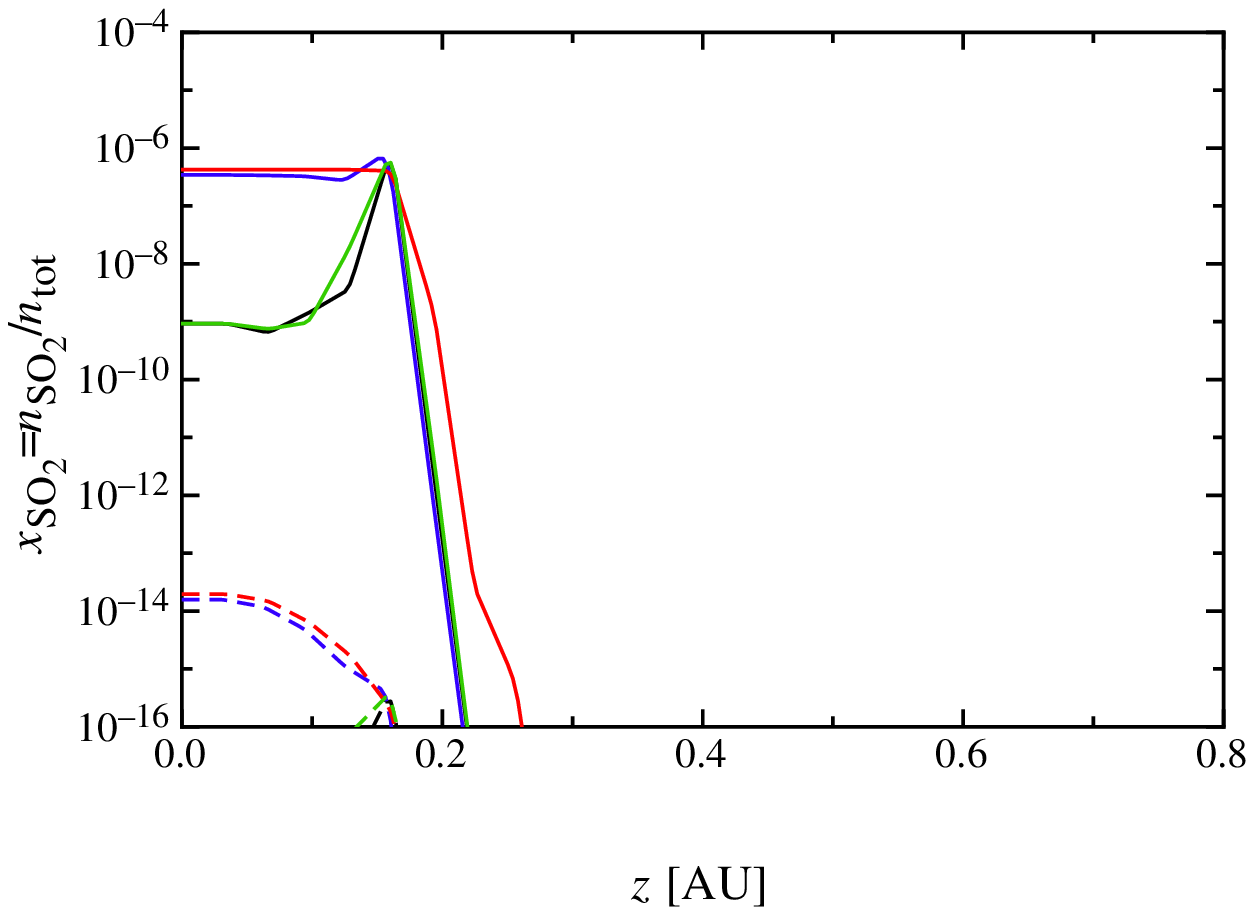}\\
\myincludegraphics{width=0.42\textwidth}{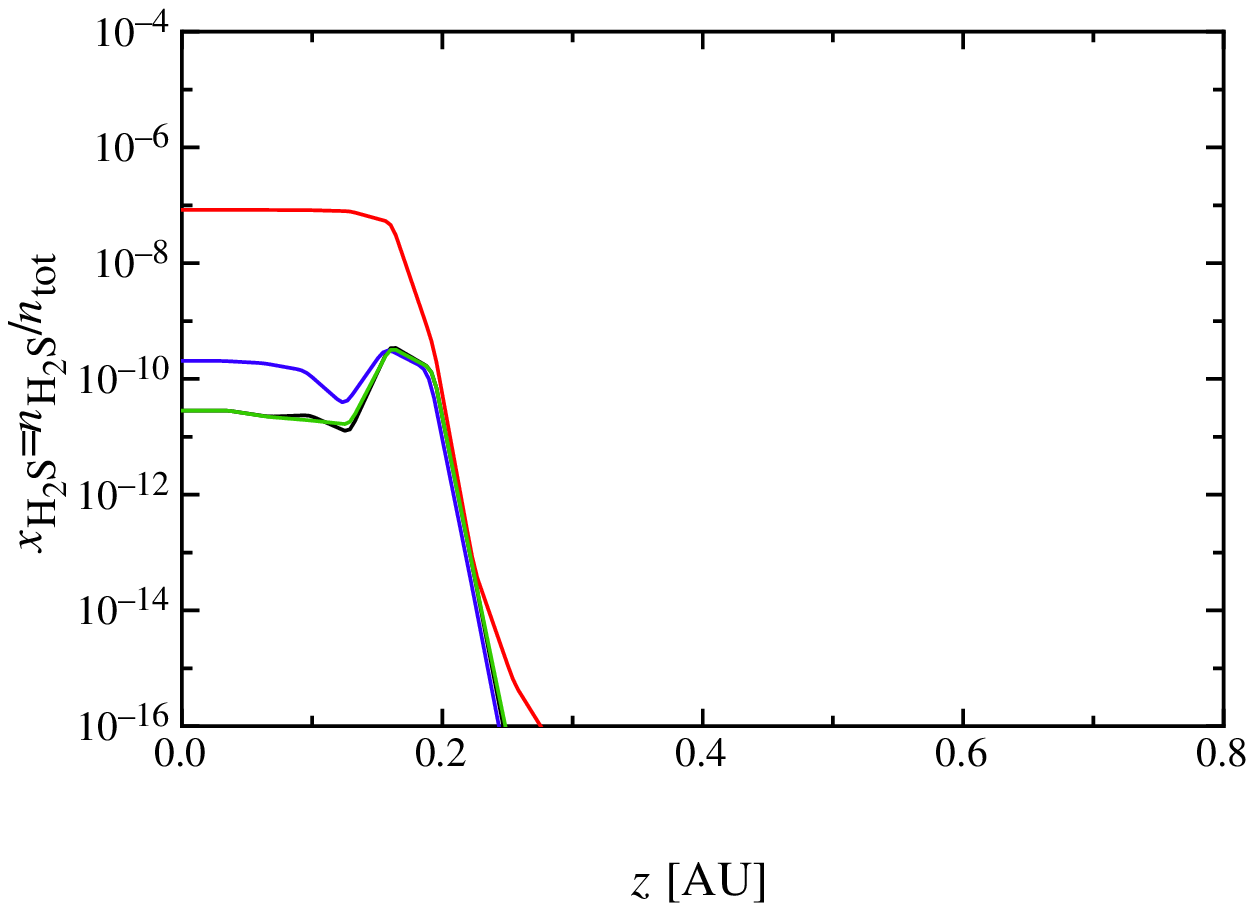}%
\caption{(cont.) Gas and ice fract. abundances at $s=1.3\,\mathrm{AU}$.}
\end{figure}

Turbulent mixing has \reftext{important effects}
on the disk's chemical composition\reftext{, in particular in the upper layers}. Generally, diffusive mixing tries to
homogenize the fractional abundances of molecules over the depth of the disk. 
Figure~\ref{fig_interesting_species_at_is18_part_1}
illustrates this effect, for example in the case of H$_2$, the fractional abundances are slightly
increased in the transition region, with only negligible changes in the column density (c.\,f., Table~\ref{tab_column_densities}
at the end of this section).

This effect is more pronounced for CO$_2$: in model A,  CO$_2$ is abundant close to the midplane only, since
it is removed from the upper layers by reactions with H and H$^+$ and by photo-dissociation due to the 
incident UV and X-ray radiation:
\pagebreak
\begin{eqnarray*}
\mathrm{H} + \mathrm{CO}_2 &\ \longrightarrow\ & \mathrm{CO} + \mathrm{OH}\,\\
\mathrm{H}^{+} + \mathrm{CO}_2 &\ \longrightarrow\ & \mathrm{HCO}^{+} + \mathrm{O}\,\\
\mathrm{CO}_{2} + \gamma_\mathrm{UV} &\ \longrightarrow\ & \mathrm{CO} + \mathrm{O}\,\\
\mathrm{CO}_{2} + \gamma_\mathrm{X-ray} &\ \longrightarrow\ & \mathrm{CO} + \mathrm{O}\,.
\end{eqnarray*}
At $z\approx 0.2\,\mathrm{AU}$, the reactions with H$^+$ and the X-ray photochemistry are dominant, while in the upper
layers, the UV photochemistry is most important. In model MIX, CO$_2$ diffuses upwards and its abundance is 
reduced near the midplane.

\reftext{The H$_2$O, OH, and CO} abundances are correlated 
with that of molecular hydrogen \reftext{and are enhanced in model MIX in the region where H$_2$ is more abundant}.
\reftext{The effect of diffusion of H$_2$O leads to reduced fractional abundances for 
$z< 0.15\,\mathrm{AU}$ and} increased abundances between $0.15\,\mathrm{AU}$ and $0.2\,\mathrm{AU}$. 
Water is destroyed easily at high temperatures and through photodissociation by UV and X-ray photons. 
The H$_2$O abundance has a dip at $z\approx 0.22\,\mathrm{AU}$, where both the gas temperature and
density are relatively low \citep[c.\,f.,][Figure~1]{woitke09b}. Here, UV and X-ray \reftext{photodestruction}
become strong, while at the same time the temperature is still too low to activate the main formation reactions for H$_2$O,
\begin{eqnarray*}
\mathrm{H}_2 + \mathrm{O} &\ \longrightarrow\ & \mathrm{OH} + \mathrm{H}\,\\
\mathrm{OH} + \mathrm{H}_2 &\ \longrightarrow\ & \mathrm{H}_2\mathrm{O} + \mathrm{H}\,.
\end{eqnarray*}

\reftext{In the case of methane, diffusion also increases the fractional abundances in the surface layers.} At the same time,
the formation of CH$_4$ close to the midplane is efficient enough to maintain its initial fractional abundance\reftext{,}
\reftext{leading} to an overall increase in the column density of methane\reftext{.}
For $z\geq 0.3\,\mathrm{AU}$, methane is destroyed quickly by UV and X-ray photons.

A similar mechanism is responsible for the considerable increase in ammonia for $z\leq 0.3\,\mathrm{AU}$\reftext{.}
The formation of ammonia near the disk midplane is driven by a change in the oxygen chemistry, 
with O$_2$ being overabundant
in model MIX (see also Section~\ref{sec_results_interaction_2}). The differences
in the oxygen abundance between model A and model MIX is caused by the continuous diffusion of
O$_2$ from the warm molecular layer towards the midplane.
This implies a decrease in the column density of HCN, whose formation is competing with NH$_3$ for the 
available nitrogen. Since the number of nitrogen
atoms over the full column is conserved and since the formation of NH$_3$ is more efficient,
HCN is depleted by one order of magnitude close to the {midplane}, compared with {that in} model A. 
Its column density decreases by 
\reftext{approximately} the amount required for the formation of the additional ammonia molecules.

The most pronounced effect arises in the case of methanol as a consequence of the increase in ammonia.
Its fractional abundance increases for
$z \leq 0.3\,\mathrm{AU}$ by one to six orders of magnitude, resulting in a column density
of $1.2 \cdot 10^{19}\mathrm{cm}^{-2}$ compared with $4.7 \cdot 10^{17}\mathrm{cm}^{-2}$ {in} model A. 
This is due to an enhanced production of CH$_3$OH by the reaction \citep{rodgers01}
\begin{eqnarray*}
\mathrm{CH}_3\mathrm{OH}_2^+ + \mathrm{NH}_3 &\ \longrightarrow\ & \mathrm{NH}_4^+ + \mathrm{CH}_3\mathrm{OH}\,.
\end{eqnarray*}
In Figure~\ref{fig_interesting_species_at_is18_part_1}, we display the ammonia and methanol 
abundances for an additional model calculation with diffusive mixing, but where this reaction
is switched off. The ammonia abundances remain unchanged, but the methanol abundances drop by one
order of magnitude, which is comparable to the midplane abundances in model~A.

Significant changes are also found for acetylene, C$_2$H$_2$. At $z=0.2\,\mathrm{AU}$, 
it is mainly produced from NH$_3$ by the reactions
\begin{eqnarray*}
\mathrm{NH}_3 + \mathrm{C}_2\mathrm{H}_3^+ &\ \longrightarrow\ & \mathrm{C}_2\mathrm{H}_2 + \mathrm{NH}_4^+\\
\mathrm{NH}_3 + \mathrm{C}_2\mathrm{H}_2^+ &\ \longrightarrow\ & \mathrm{C}_2\mathrm{H}_2 + \mathrm{NH}_3^+\,,
\end{eqnarray*}
as well as by a reaction of C and CH$_3$ at $z<0.2\,\mathrm{AU}$,
\begin{eqnarray*}
\mathrm{C} + \mathrm{C}\mathrm{H}_3 &\ \longrightarrow\ & \mathrm{C}_2\mathrm{H}_2 + \mathrm{H}\,.
\end{eqnarray*}
Switching off these reactions leads to a significant decrease in C$_2$H$_2$ 
(c.\,f., Figure~\ref{fig_interesting_species_at_is18_part_1}).
Hence, acetylene profits from the increase in ammonia abundances and is 
overabundant by two orders of magnitude at $z\leq 0.2\,\mathrm{AU}$. Higher up in the disk,
it is destroyed efficiently by photochemical reactions and reactions with atomic hydrogen.

Finally, we inspect the sulphur-bearing species SO$_2$ and H$_2$S.
Initially, sulphur is locked in H$_2$S only, with a fractional
abundance of $10^{-6}$ (c.\,f., Table~\ref{tab_umist_initial_abundances}). 
H$_2$S is easily destroyed by atomic hydrogen due to the small activation barrier of this reaction. 
Provided the O$_2$ and OH abundances are sufficiently high, this drives \reftext{a} chain of 
reactions \citep{charnley97,millar97}
\reftext{which forces large amounts of H$_2$S into SO$_2$ in both models at $z \leq 0.25\,\mathrm{AU}$.}
The now highly abundant SO$_2$ near the midplane is destroyed on timescales longer than its formation timescales
and therefore provides a steady reservoir to maintain the enhanced H$_2$S abundances in comparison to model A. 
At the midplane, the destruction
of H$_2$S is less efficient, but still sufficient to produce OCS from CO and S \citep{hatchell98}. 
After $10^6\mathrm{yr}$, sulphur is locked in SO$_2$ and OCS in equal amounts.

Contrarily, in model A, the formation of SO$_2$ is inefficient in the lower layers due to the missing OH and the missing
O$_2$ \reftext{at $z \approx 0$}. Thus, the slower reaction path of S with CO is preferred and
OCS is built up over the lifetime of the disk, eventually locking all sulphur.

\subsubsection{Disk winds}
The resulting chemical abundances for model WND are displayed in 
Figure~\ref{fig_interesting_species_at_is18_part_1}.
For all species, the effects of the disk wind on the vertical stratification 
of the chemical species are much less pronounced than in the case of turbulent mixing.
In the cases of CO$_2$, NH$_3$, CH$_3$OH and C$_2$H$_2$, slight increases in the 
fractional abundances are found around $z=0.2\,\mathrm{AU}$, in particular where narrow
dips occur in the abundances from model A. Protected by the upper layers against the 
\reftext{UV} and X-ray irradiation, molecules transported upwards from the midplane
fill these sinks on timescales comparable to their removal by chemical reactions.

The reason for the inefficiency of the disk wind lies in the steady disk model assumption
considered here: to avoid disk dispersal over the lifetime of
the disk, which would contradict the steady state assumption, we introduced a scaling factor 
$\varkappa = 10^{-3}$ for the resulting wind speeds of \citet{suzuki09}
at the midplane layer. Hence, the upward velocities near the midplane are 
low in model WND, compared to the turbulent velocities in model MIX
(c.\,f., Figure~\ref{fig_timescales_velocities}). At radii $s\leq 10\,\mathrm{AU}$, 
the wind velocities become very large in the \reftext{photodestruction} region of
the disk ($z/s>0.3$), due to the continuity equation in vertical direction in steady state.
But there, the timescales of the \reftext{photochemistry} are extremely short so that the effect of 
the disk wind is small (c.\,f., Figure~\ref{fig_timescales_velocities}).

For a scaling factor $\varkappa = 10^{-3}$, the midplane grid cell at radius $s=1.3\,\mathrm{AU}$
evaporates in about $2 \cdot 10^5\mathrm{yr}$, corresponding to a mass loss of $20\%$ of the total mass contained
in this disk ring. Hence, in a stationary model, $\varkappa$ should be lower than $10^{-3}$, which
would lead to even less efficient disk winds.

We thus conclude that in a steady disk model, the disk winds have only negligible effects. 
Due to the necessary downscaling of
the wind speeds by a factor of $1000$ ($\varkappa=10^{-3}$), the vertical mass transport 
becomes small and causes only small changes in the disk's chemical composition.
Nonetheless, it is worth noting that in a time-dependent physical disk model, 
where evaporation and dynamical refueling by the accretion flow
are allowed, the disk wind may become significant. The simple model considered here suggests that 
it will affect molecules that also show an enhancement in
the upper layers in the diffusive mixing model (e.\,g., NH$_3$, CH$_3$OH).

\subsubsection{Viscous accretion}
The chemical abundances from model ACR at a radius of $1.3\,\mathrm{AU}$ 
are displayed in Figure~\ref{fig_interesting_species_at_is18_part_1}.
As in the previous cases, at the disk surface photochemistry dominates and the
accretion flow does not affect the chemical composition. Close to the disk midplane, 
the accretion time scale is comparable to the chemical
timescale and the effects of the radial inflow of material become visible. 
Whether the accretion flow leads to an increase or a decrease in the abundance of a particular species depends
on the formation and destruction timescales of that species. 
In the case of H$_2$O, for example, the chemical reactions are fast enough to compensate for the effects of the
accretion flow and the abundances and column density are similar to those from model A. 

On the contrary, NH$_3$ shows higher abundances at $z<0.15\,\mathrm{AU}$.
As discussed for the case of diffusive mixing above, the total amount of nitrogen atoms is conserved and determined
by the initial abundances of N$_2$ and NH$_3$ (c.\,f., Table~\ref{tab_umist_initial_abundances}). 
An increase in the column density of NH$_3$ from
$2.5 \cdot 10^{20}\mathrm{cm}^{-2}$ in model A to $3.4 \cdot 10^{20}\mathrm{cm}^{-2}$ in model ACR
leads to a decrease in HCN from $4.3 \cdot 10^{19}\mathrm{cm}^{-2}$ to $1.1 \cdot 10^{19}\mathrm{cm}^{-2}$ at this radius.

In contrast to the results from model MIX, methanol shows lower abundances close to the disk midplane, 
despite the increase in ammonia. Since its fractional
abundance is even lower than those from model A, the reason, therefore, 
is not only the lack of vertical diffusion, 
but the fact that the formation and evaporation timescales (from dust grains) of methanol 
at around $10\,\mathrm{AU}$ are longer than its destruction timescale (c.\,f., Section~\ref{sec_results_interaction_2}).
The accretion flow propagates this depletion inwards over time. Compared to model A, this reduces the CH$_3$OH 
column density from $4.7 \cdot 10^{17}\mathrm{cm}^{-2}$ to $1.4 \cdot 10^{16}\mathrm{cm}^{-2}$.

One should bear in mind that the location of the snow line for methanol and hence 
its gas-phase abundance is sensitive to its binding energy
on dust grains, which varies across the literature. Here, we adopt the theoretical value from \citet{hasegawa93}, 
$E_\mathrm{bind}=2140\,\mathrm{K}$.
\citet{sandford93} and more recently \citet{brown07} find much higher values between $4000$ and $5000\,\mathrm{K}$ 
based on the results of temperature-programmed desorption (TPD) experiments on pure methanol ice. Higher binding
energies would imply lower gas-phase abundances and a shift of the snow line to smaller radii.
Here, the NH$_3$ abundances are higher and the formation of methanol more efficient, 
which would imply higher gas-phase abundances within the snow line.

A similar decrease in the abundances in the innermost disk can be found particularly for 
C$_2$H$_2$, where the accretion flow removes a considerable fraction of acetylene close to the midplane 
(c.\,f., Table~\ref{tab_column_densities} for the resulting column densities of C$_2$H$_2$ and other species).
\begin{deluxetable}{rlllll}
\tabletypesize{\scriptsize}
\tablecaption{Calculated column densities $N [\mathrm{cm}^{-2}]$ at radius $s=1.3\,\mathrm{AU}$
of selected species for the disk models considered in this study.\label{tab_column_densities}}
\tablewidth{0pt}
\tablehead{\colhead{species} & \colhead{A} & \colhead{B} & \colhead{ACR} & \colhead{MIX} & \colhead{WND}}
\startdata
H$_2$      & 1.7e+25 & 1.7e+25 & 1.7e+25 & 1.7e+25 & 1.7e+25\\
CO$_2$     & 2.5e+19 & 2.6e+19 & 2.5e+19 & 8.3e+18 & 2.4e+19\\
H$_2$O     & 4.2e+20 & 4.2e+20 & 4.0e+20 & 3.5e+20 & 4.2e+20\\
OH         & 1.5e+15 & 9.3e+15 & 1.5e+15 & 5.3e+15 & 1.8e+15\\
CO         & 1.6e+21 & 1.6e+21 & 1.6e+21 & 1.7e+21 & 1.6e+21\\
CH$_4$     & 5.4e+18 & 5.4e+18 & 6.5e+17 & 7.1e+18 & 5.7e+18\\
NH$_3$     & 2.5e+20 & 2.5e+20 & 3.4e+20 & 3.0e+20 & 2.6e+20\\
HCN        & 4.3e+19 & 3.8e+19 & 1.1e+19 & 4.7e+18 & 4.0e+19\\
CH$_3$OH   & 4.7e+17 & 4.8e+17 & 1.4e+16 & 1.2e+19 & 5.6e+17\\
C$_2$H$_2$ & 1.6e+16 & 1.6e+16 & 1.3e+14 & 2.2e+18 & 1.6e+16\\
SO$_2$     & 2.7e+17 & 2.6e+17 & 1.2e+19 & 1.4e+19 & 2.9e+17\\
H$_2$S     & 9.6e+14 & 8.3e+17 & 6.2e+15 & 2.8e+18 & 9.4e+14\\
OCS        & 3.3e+19 & 3.2e+19 & 2.0e+19 & 1.6e+19 & 3.3e+19
\enddata
\end{deluxetable}
\pagebreak

\subsection{Accretion, mixing -- full disk solutions}\label{sec_results_interaction_2}
In the previous section, we inspected in detail the chemical abundances for the different disk models 
as a function of height at a single radial position. The chemical
composition varies between the individual models due to mass transport in the radial and vertical directions. 
The results thereby depend
strongly on the species considered, in particular, on its formation and destruction timescales relative to 
the dynamical timescales.
Since these timescales vary with radial distance from the star for accretion, turbulent mixing and disk winds, 
we expect significant changes in
the effects of these mass transport phenomena at different radial positions in the disk.

Here, we focus on the radial accretion and vertical mixing and neglect the disk wind. 
At a radius, $s=1.3\,\mathrm{AU}$, the influence of the disk wind on the chemistry is small, compared with diffusive mixing, 
due to the low wind speed required to impede the evaporation of the disk in our model. At larger radii, the wind velocities
are even smaller, leading to longer timescales of the upward wind motion. 
At the same time, however, the chemical timescales
increase 
(c.\,f., Section~\ref{sec_timescales}), so that we do not find significant changes 
to the disk chemistry due to the inclusion of a disk wind.

\begin{figure*}
\myincludegraphics{width=0.9\textwidth}{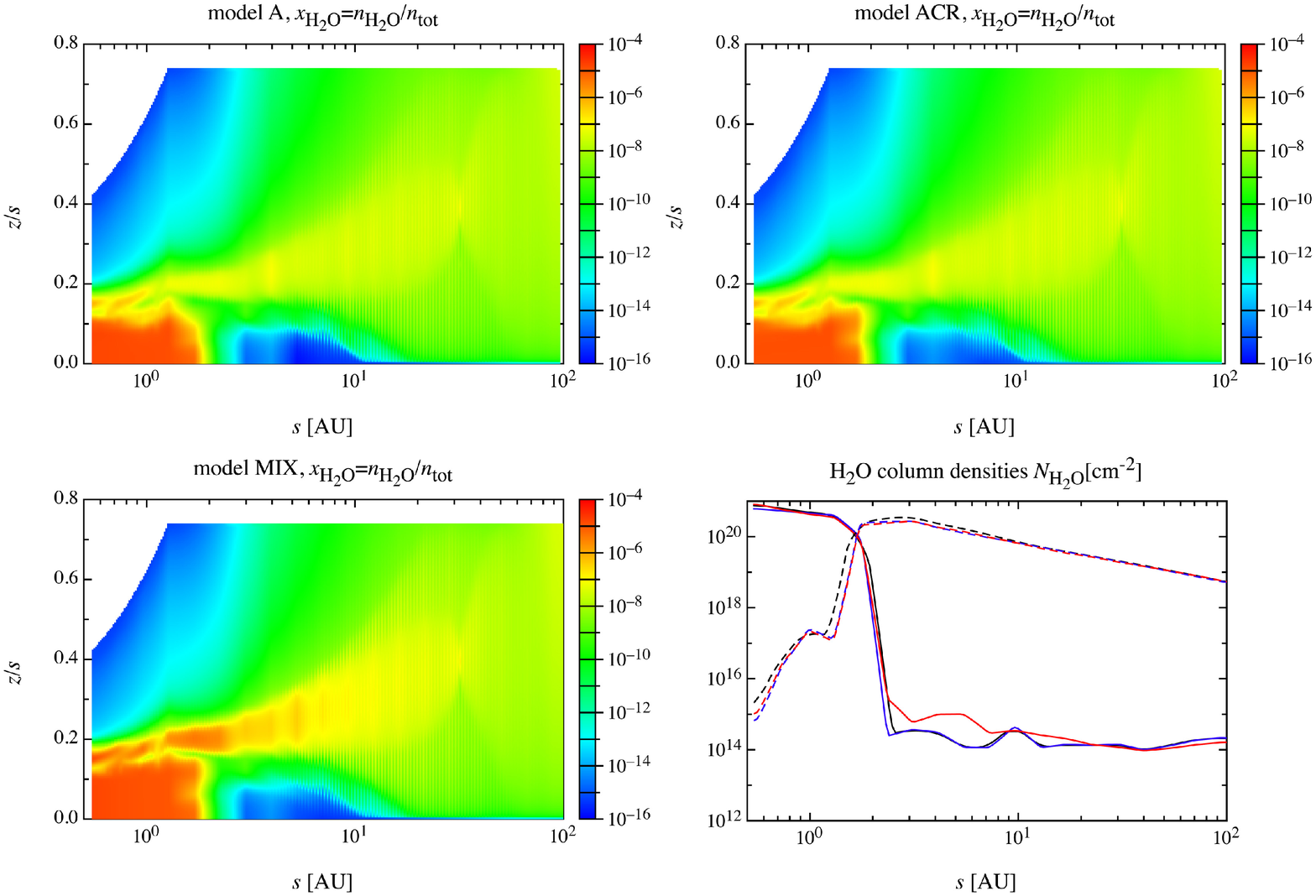}\\
\myincludegraphics{width=0.9\textwidth}{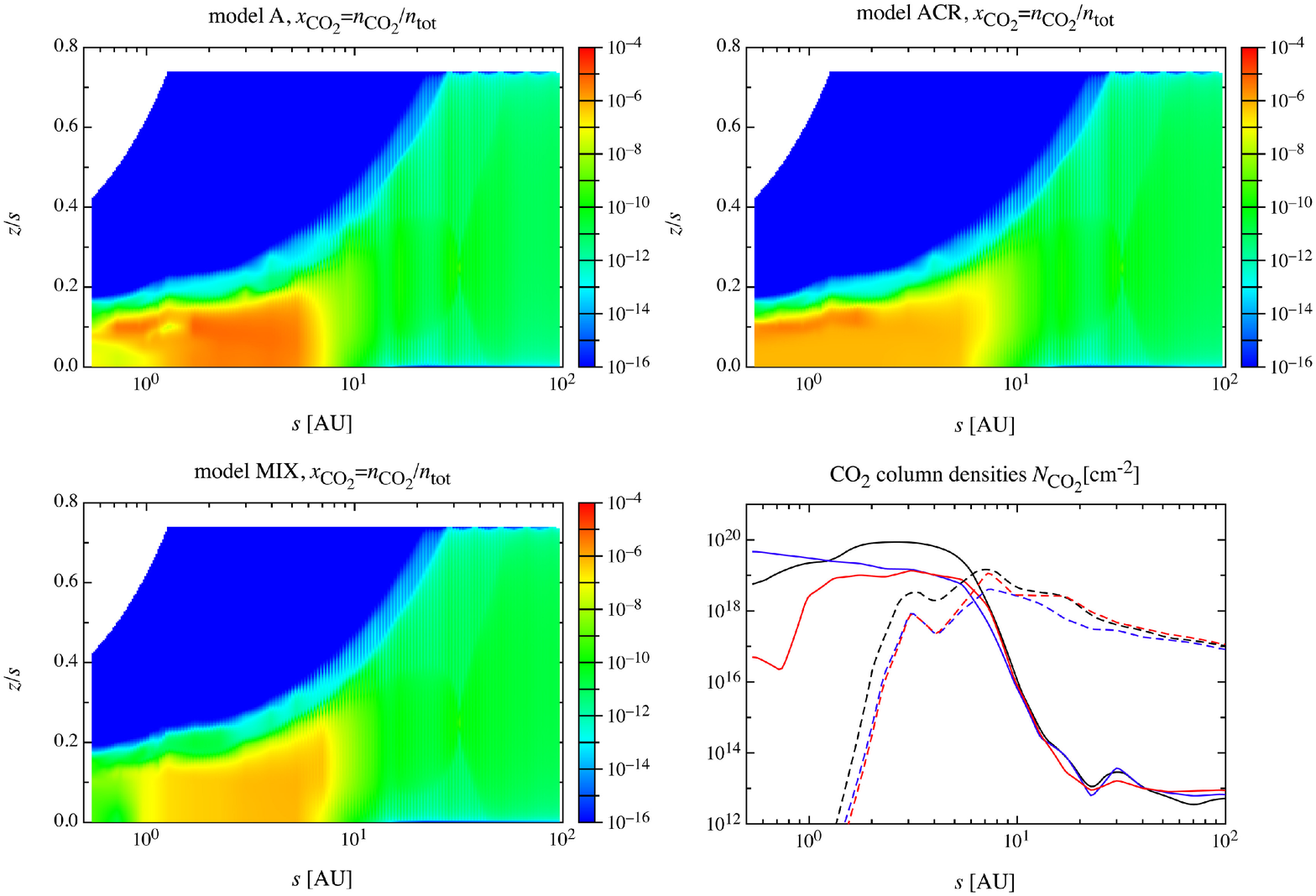}%
\caption{Two-dimensional fractional gas-phase abundances and column densities (solid: gas; dashed: ice)  of H$_2$O and CO$_2$
for model A (black), model ACR (blue) and model MIX (red).}\label{fig_interesting_species_disk_1}%
\end{figure*}
\addtocounter{figure}{-1}
\begin{figure*}
\myincludegraphics{width=0.9\textwidth}{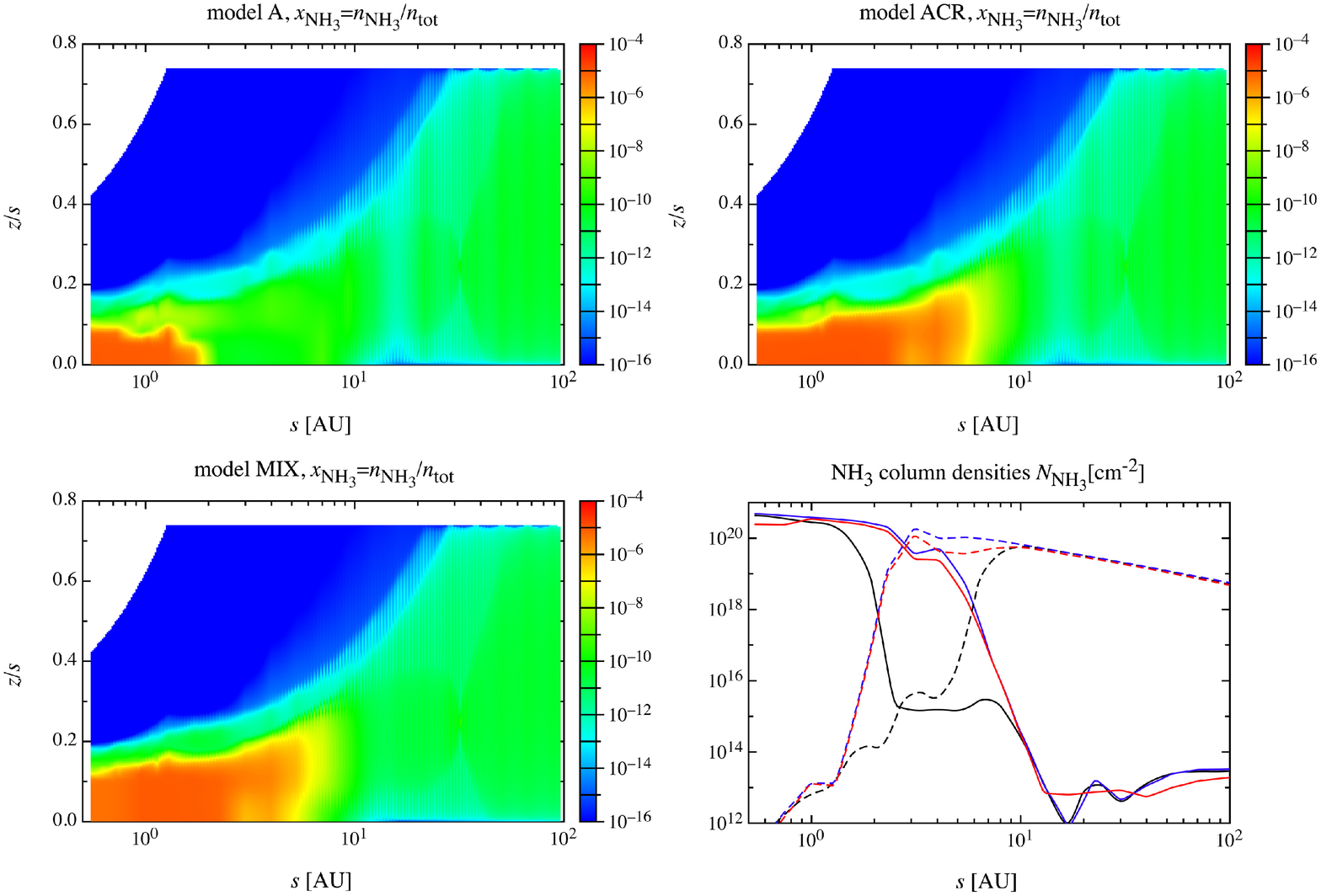}\\
\myincludegraphics{width=0.9\textwidth}{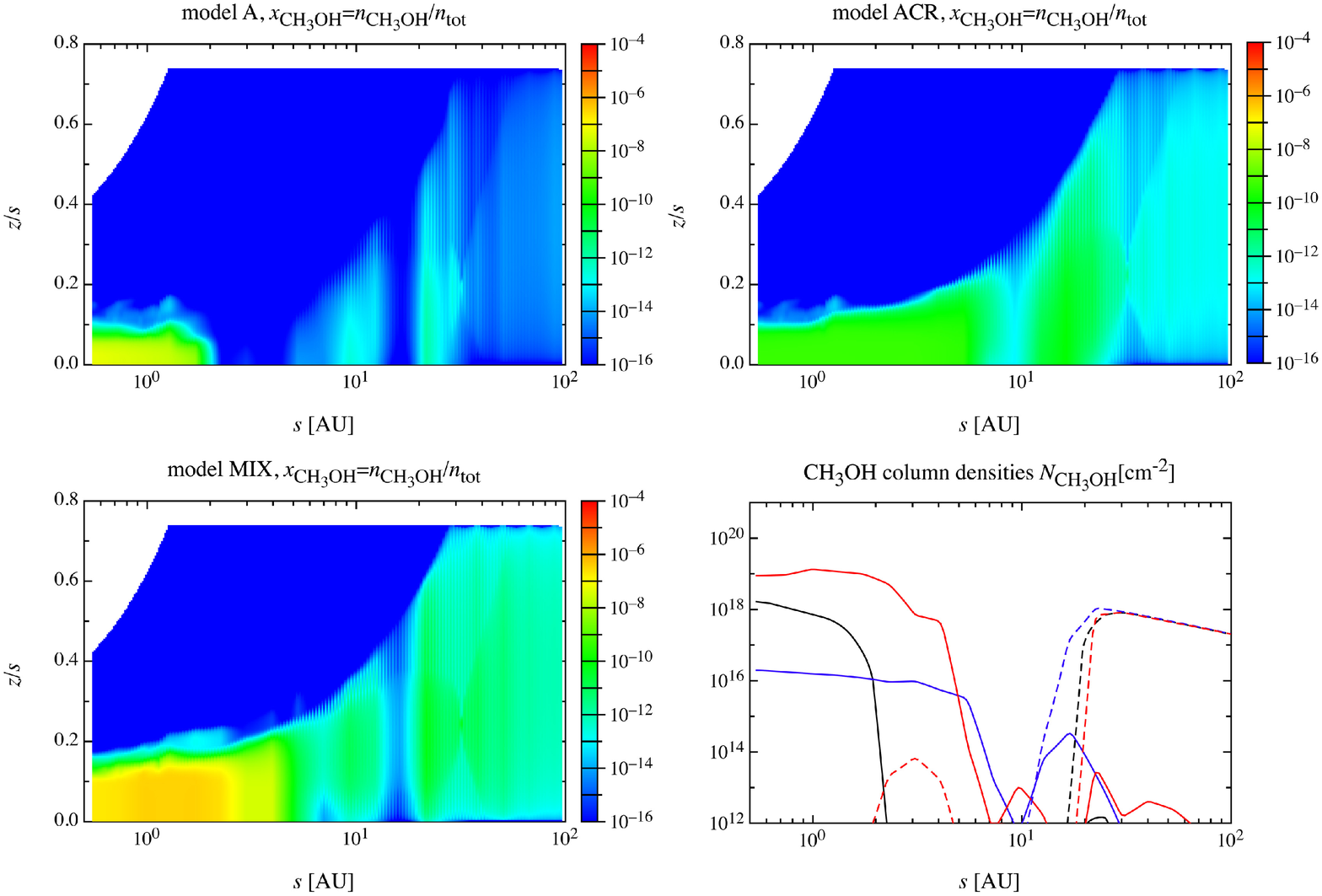}%
\caption{(cont.) Two-dimensional fractional gas-phase abundances and column densities (solid: gas; dashed: ice)  
of NH$_3$ and CH$_3$OH for model A (black), model ACR (blue) and model MIX (red).}
\end{figure*}
\addtocounter{figure}{-1}
\begin{figure*}
\myincludegraphics{width=0.9\textwidth}{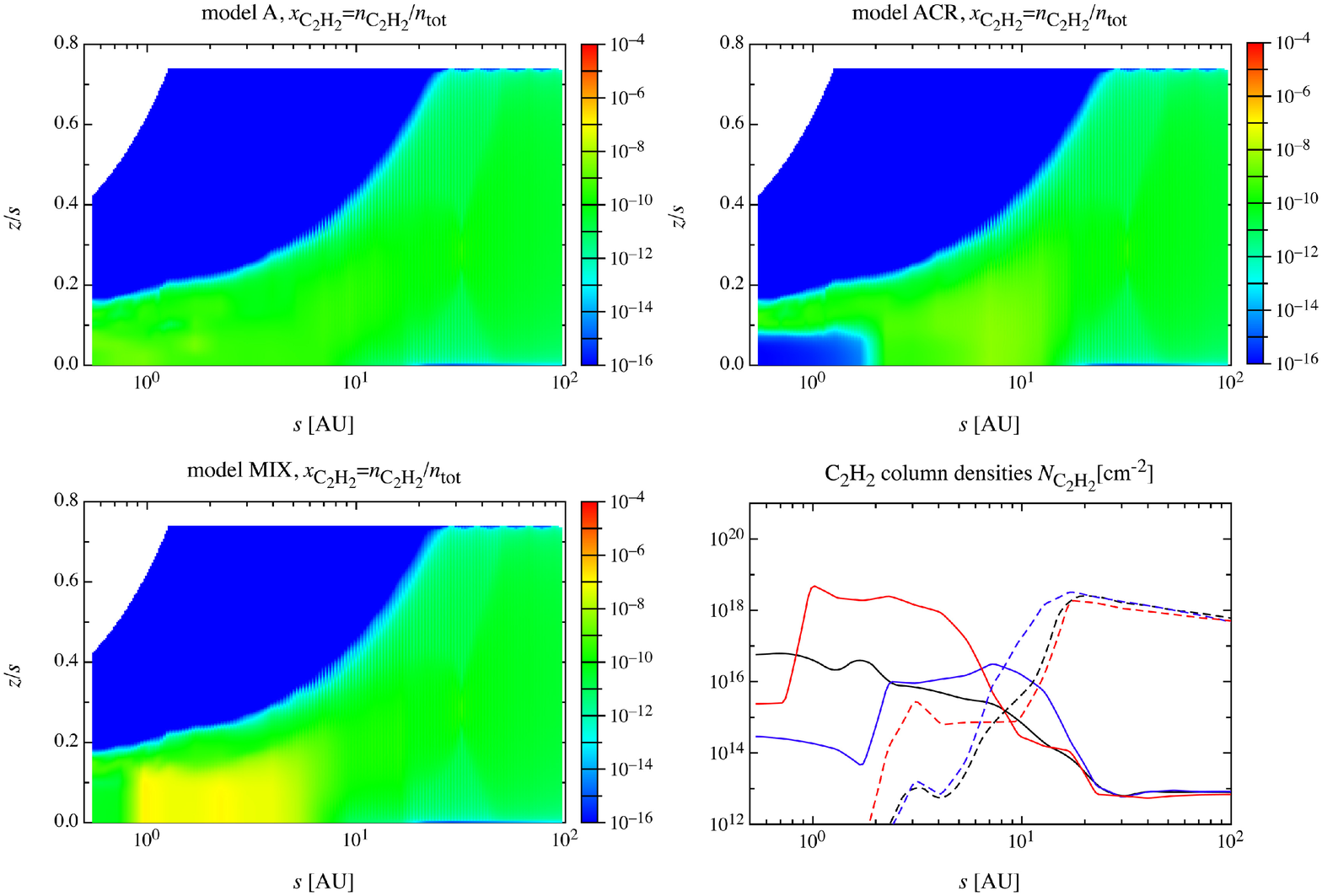}%
\caption{(cont.) Two-dimensional fractional gas-phase abundances and column densities (solid: gas; dashed: ice)  of C$_2$H$_2$
for model A (black), model ACR (blue) and model MIX (red).}
\end{figure*}

\begin{figure*}
\myincludegraphics{width=0.99\textwidth}{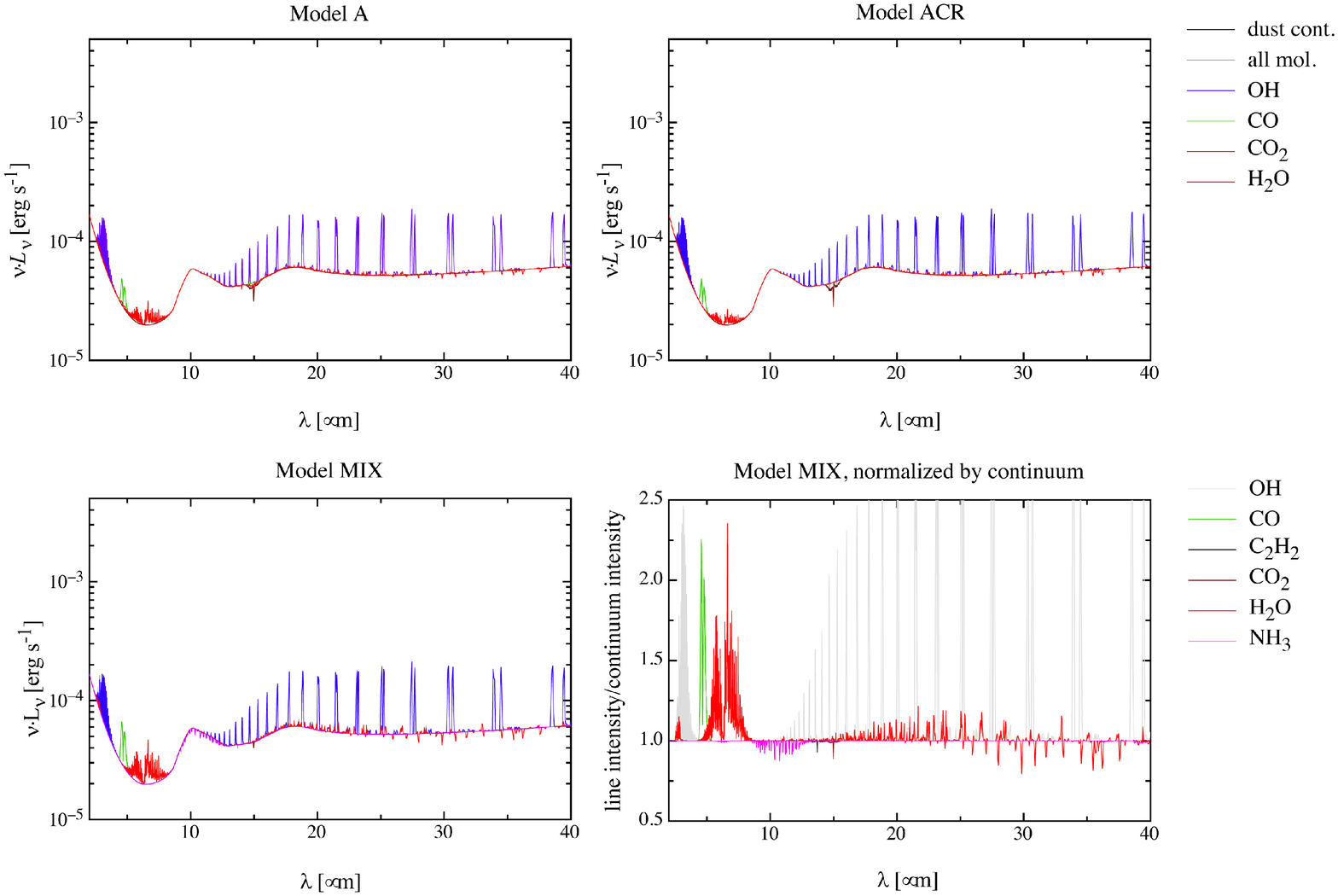}%
\caption{\reftext{LTE} line spectra for model A, model ACR and model MIX for an observer at inclination $i=0$. 
For model MIX, we also show the spectrum normalized by the continuum flux in the lower right panel
(for a better illustration, the color coding in the normalized spectrum is different from the other
spectra and listed in lower right legend).}\label{fig_spectra_different_models}
\end{figure*}

We compare the resulting fractional gas-phase abundances and the gas and ice column densities for
model ACR and model MIX with model A in Figure~\ref{fig_interesting_species_disk_1}.
The results for water differ only slightly between the models, and the
only difference arises in a narrow stripe $z/s\approx 0.2$--$0.4$, where the gas-phase abundances 
of water are increased in model MIX  due to the enhanced H$_2$ abundances (see also the discussion in
Section~\ref{sec_grain_formation}). 
This leads to an increase in the gas-phase column density between $2$ and $10\,\mathrm{AU}$. 
Since the H$_2$O abundances are increased in the transition layer,
turbulent mixing affects the strength of the H$_2$O line emission from the disk. 
This can be seen in Figure~\ref{fig_spectra_different_models},
where we compare the emerging disk spectra for model A, model ACR and model MIX. 
Apart from H$_2$O, the species dominating the molecular line emission
are CO and OH. Both show slightly stronger emission lines in model MIX and in model ACR,
although the effect is more pronounced in the case of diffusive mixing.

In the case of CO$_2$, turbulent mixing reduces the gas-phase column density
inside $10\,\mathrm{AU}$: \reftext{turbulent} mixing leads
to the diffusion of CO$_2$ from the lower, protected layers into the upper disk, 
where it is efficiently destroyed by incident radiation, compared
with its production. For radii $\leq 1\,\mathrm{AU}$, CO$_2$ is strongly depleted. 
Viscous accretion, on the other hand, enhances the CO$_2$ gas-phase abundances
at radii $\leq 1\,\mathrm{AU}$ due to a net inward transport from radii $\leq 10\,\mathrm{AU}$. 
This leads to changes in the disk spectra for both models ACR and MIX: inside $2\,\mathrm{AU}$, 
the gas and dust temperature first decrease with increasing distance from the
midplane (see Figure~\ref{fig_disk_properties}). 
This is due to the viscous heating near the midplane.
Since CO$_2$ is concentrated near the midplane and since it is abundant in
model ACR (depleted in model MIX), the absorption feature at $15\micron$ is enhanced (reduced).

Ammonia shows the largest variations between the different models. 
Compared to model A, model ACR and model MIX lead to a similar increase
in the gas and ice abundances between $1$ and $10\,\mathrm{AU}$. 
The increase in ammonia is a result of a fundamental change in the oxygen chemistry. In
model ACR and model MIX, atomic oxygen is reduced by orders of 
magnitude between $2$ and $10\,\mathrm{AU}$, with an increase in O$_2$
and a decrease in OH. The drastic decrease in atomic oxygen practically 
switches off the the main destruction reaction of NH$_3$,
\begin{eqnarray*}
\mathrm{O} + \mathrm{NH}_3 &\ \longrightarrow\ & \mathrm{OH} + \mathrm{NH}_2\,.
\end{eqnarray*}
The overabundant regions are located below the warm molecular layer in model
ACR and reach slightly into it in model MIX (see also Figure~\ref{fig_interesting_species_at_is18_part_1}). 
Hence, ammonia is found in absorption
in the spectrum of model MIX due to the same reasons as CO$_2$ in model ACR.

Another species of interest is methanol, whose gas-phase abundance is high in regions
where the ammonia abundances are high in all the models. In model MIX, this implies an
increase in methanol within $5\,\mathrm{AU}$ compared to model A. Viscous accretion
increases its abundance between $2$ and $5\,\mathrm{AU}$ and decreases it further inside. 
As in model MIX, CH$_3$OH is produced
efficiently from NH$_3$, however, its gas-phase formation and evaporation from dust grains at 
$5$--$10\,\mathrm{AU}$ is slower than its destruction
and the accretion flow transports less material into the inner disk than is removed from it.
Methanol is located below the transition region of the disk, even in model MIX, hence, 
the line opacities are so small that the absorption features are negligible in the spectra.

The abundance profile of acetylene can be explained as follows: 
\reftext{in} model MIX, its \reftext{abundance is} increased between $1$ and $10\,\mathrm{AU}$.
Further inside, the diffusion leads to a depletion of C$_2$H$_2$. 
In model ACR, C$_2$H$_2$ is increased between $2$ and $10\,\mathrm{AU}$ and decreased
further inside, similar to CO$_2$. Since the overall abundance of C$_2$H$_2$ is low and since it is 
located below the transition region,
the absorption lines are negligibly small in the spectra of the disks in model A and model ACR.
The stark increase in C$_2$H$_2$ in model MIX leads to a weak absorption feature seen at $13$--$13.5\micron$
(Figure~\ref{fig_spectra_different_models}).

\subsection{Theory {versus} observations}\label{sec_discussion_observations}
We refer to the observations of the protoplanetary disks of AA\,Tauri \citep{carr08}, 
DR\,Tauri and AS\,205 \citep{salyk08}.
AA\,Tauri is a classical T\,Tauri star with mass $0.8 M_\odot$, accretion rate 
$10^{-8} M_\odot \mathrm{yr}^{-1}$ and inclination $75^\circ$ \citep{bouvier99}.
\citet{carr08} observed the disk around AA\,Tauri with the Infrared Spectrograph (IRS) 
on the Spitzer Space Telescope over the wavelength
range of $9.9$--$37.2\micron$. They found a rich spectrum of molecular emission lines 
dominated by rotational transitions of H$_2$O and
OH. Between $10$ and $15\micron$, they detected ro-vibrational emission bands of C$_2$H$_2$, 
HCN and CO$_2$, plus the atomic [Ne\,II] transition
at $12.8\micron$. Fundamental ro-vibrational emission bands of CO have been detected in 
Keck/NIRSPEC observations between
$3$ and $5\micron$. Fitting model spectra to the observations, \citet{carr08} derive a high column density for
these species and characteristic emission radii in the inner disk ($s<2.5\,\mathrm{AU}$).

DR\,Tauri and AS\,205, both classical T\,Tauri stars, have been observed with IRS on Spitzer and 
with Keck/NIRSPEC by \citet{salyk08}. DR\,Tauri has a mass of
$0.76 M_\odot$ and a variable accretion rate of $[0.3$--$79]\cdot 10^{-7} M_\odot \mathrm{yr}^{-1}$, 
much higher than in our model or that observed in AA\,Tauri.
AS\,205 is the primary component of a triple system with mass $1.2 M_\odot$ and accretion 
rate $7.2\cdot 10^{-7} M_\odot \mathrm{yr}^{-1}$.
The inclinations are $67^\circ$ for DR\,Tauri and $47^\circ$ for AS\,205A, respectively. 
Both systems show a large number of transitional emission
lines from water and require high temperatures and column densities to explain this emission, 
similar to AA\,Tauri. 
Further, CO, OH and CO$_2$ are required to fit the observed
spectra in the near- to mid-infrared. In the following, we discuss how our models 
relate to, and compare with, these observational results.

The calculated spectra for our models show emission and absorption features different 
from the observed spectra (Figure~\ref{fig_spectra_different_models}). Firstly, the molecular line emission
in our models is dominated by rotational transitions of OH. Weak emission from H$_2$O 
is detected in model MIX. In model ACR and in model A, water is found mainly in absorption.
Model WND leads to results similar to model A due to the limitations of our disk model. 
\reftext{These models make use of the simple H$_2$ formation rates on grains,
Eq.~\eqref{eqn_h2_grain_formation_model_A}.} Model B, 
\reftext{which accounts for the enhanced H$_2$ formation rates on grains, Eq.~\eqref{eqn_h2_grain_formation_model_B},}
leads to huge emission lines from OH and H$_2$O. The near-infrared emission bands of CO at
$5\micron$ are reproduced in model A, model MIX and also in model ACR.

\begin{figure}
\myincludegraphics{width=0.99\columnwidth}{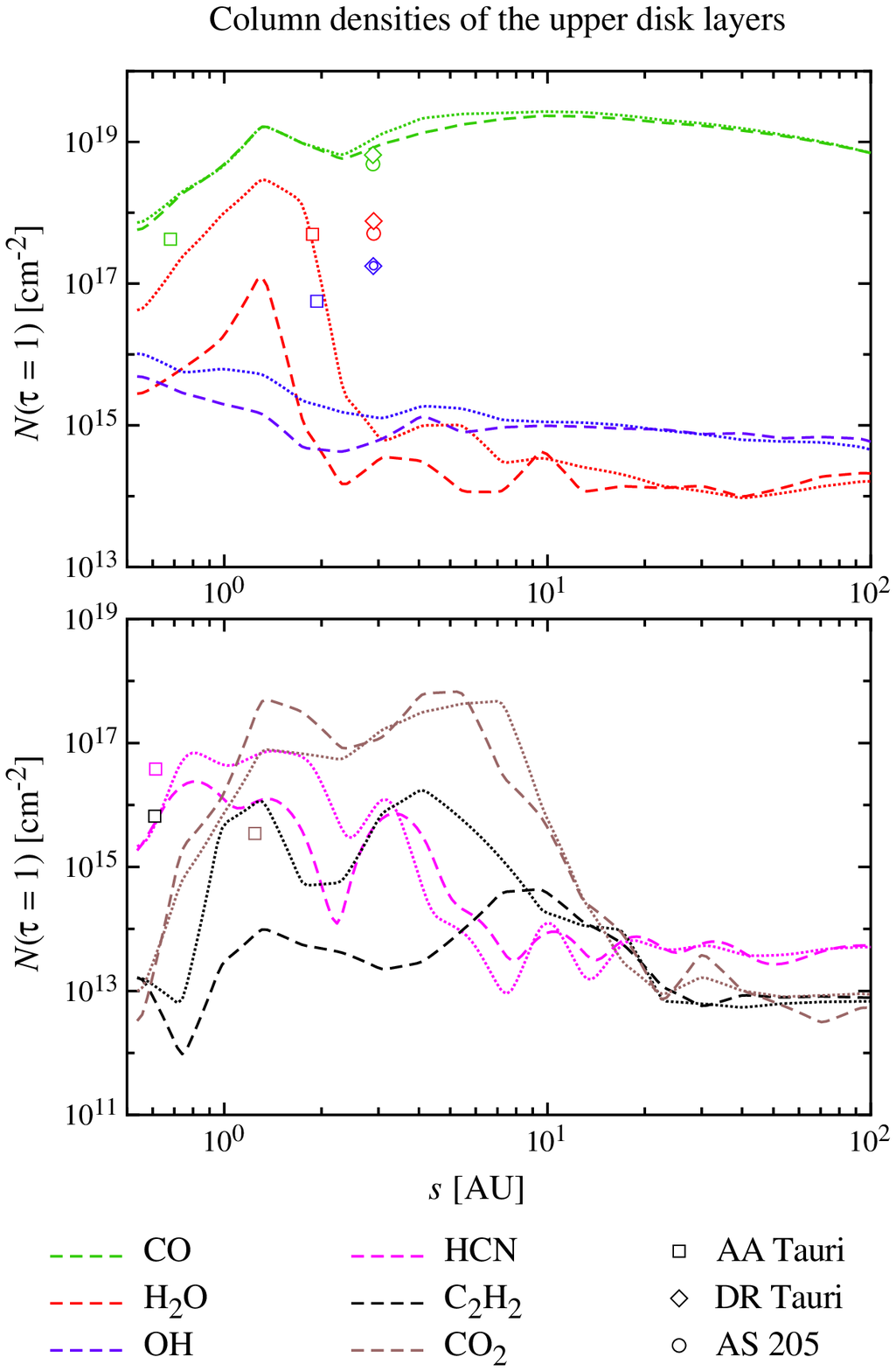}%
\caption{Column densities of the gas-phase species in the upper layers in model A (dashed)
and model MIX (dotted). The derived column densities for AA\,Tauri, DR\,Tauri and AS\,205 are
indicated at their characteristic radii \citep{carr08,salyk08}.}\label{fig_column_densities_upper_layers}
\end{figure}

The modeled line intensities depend strongly on the gas temperature and molecular 
abundances profiles, i.\,e., the detailed thermal balance and the chemistry.
It is therefore more favorable to compare the molecular column densities from our model
calculations to the observed ones. \reftext{In the inner disk, the observed column
densities and abundance ratios are derived from the photodestruction
layer and the warm molecular layer, i.\,e., from the disk surface down to an 
optical depth of about unity. The total molecular column densities in our model are
mainly determined by the dense midplane layers and are therefore 
larger by orders of magnitude and not comparable to the observed column densities. 
Thus, we calculate the column densities of the observed species from the disk surface down
to a continuum optical depth of unity at each wavelength
and compare them to the observations (Figure~\ref{fig_column_densities_upper_layers}).
At $s=1.3\,\mathrm{AU}$ and wavelengths $\lambda=\{5,10,20,40\}\,\micron$,
the disk becomes optically thick at $z/s=\{0.14, 0.15, 0.14, 0.11\}$, respectively.
For the unusually strong OH lines, a continuum optical depth of unity corresponds
to a maximum optical depth of $\tau\sim20$ (see discussion below for large discrepancy between the
models and the observations of OH lines). 
For all other molecules, the maximum optical depth of the lines is $\tau\sim 2$.}

\reftext{If we compare the H$_2$O and CO column densities of the upper layers
at disk radius $1.3\,\mathrm{AU}$, we find an abundance ratio of $0.02$
for model A and model ACR, while model MIX shows an abundance ratio of $0.18$.}
\citet{carr08} derive an abundance ratio of H$_2$O to CO of $1.3$, while \citet{salyk08} find water
to be less abundant than carbon monoxide with an abundance ratio \reftext{for} H$_2$O to CO of $0.1$ for both systems.
Note that \citet{carr08} calculate the abundance ratio for different radii of the H$_2$O 
and CO column densities ($2.1\,\mathrm{AU}$ and $0.7\,\mathrm{AU}$), while \citet{salyk08}
obtain their results for the same radius ($3\,\mathrm{AU}$).
It is important to note that the observed systems differ from our model in their central
objects, disk masses, inclination angles and accretion rates, with AA\,Tauri being the closest match.
Further, the derived column densities in \citet{carr08} and \citet{salyk08} 
contain considerable uncertainties in their values and their characteristic radii.

The resulting column densities of the upper disk layers are displayed in Figure~\ref{fig_column_densities_upper_layers}
for model A and for model MIX only, since it leads to the largest changes in the vertical distribution of the chemical
species. In model MIX, the H$_2$O to CO abundance ratio is of the order of $0.2$ inside the snow line, while 
in model A, it is reduced by at least a factor of $10$. Model MIX reproduces the H$_2$O abundance in AA\,Tauri at a radius 
of $2.1\,\mathrm{AU}$. In the case of DR\,Tauri and AS\,205, the central star is
more luminous than in AA\,Tauri and in our model, hence, the snow line is
located further outwards and gas-phase H$_2$O remains abundant with
derived column densities of the same order as in our model or in AA\,Tauri at 
radii of $2\,\mathrm{AU}$. The modeled CO abundances
agree with the observations for both models. Both models fail
to reproduce the high OH abundances, although the calculated line emission is 
much stronger than that of the observations. This adds further support
to the call for non-LTE spectral calculations and gas temperature profile 
calculations accounting for dust-grain settling and grain growth 
(c.\,f., Section~\ref{sec_grain_formation}).
Since the binding energy of OH on dust grains is low and considering it is robust 
against UV radiation, efficient turbulent mixing can increase the abundances in the upper layers 
of the disk (c.\,f., Figure~\ref{fig_column_densities_upper_layers}).

CO$_2$ is found in absorption instead of in emission in model A, model ACR
and model MIX. The absorption is strongest in model ACR and weakest in model MIX.
Contrary to this, the observations detect CO$_2$ 
in emission at a characteristic radius of $1.2\,\mathrm{AU}$ in AA\,Tauri.
It is also detected in emission towards DR\,Tauri and AS\,205 (with large uncertainties on the column 
densities and the emission radii).
The column densities derived from the observations of AA\,Tauri are lower than in model 
A and model MIX, although the latter one gives a better fit. 
Since CO$_2$ is destroyed quickly in the upper layers of the disk, stronger 
mixing will lead to lower CO$_2$ abundances.

HCN is detected in weak emission in AA\,Tauri and a number of other sources \citep[see, for example, ][]{pontoppidan10}, 
but not detected in our modeled spectra.
The column density in AA\,Tauri is slightly higher than in our models, and model MIX
again gives a better fit. Turbulent mixing
increases the column density in the upper layers, although the total column density is reduced. 
Stronger mixing will help to improve the fit of the model calculations to the observations.

In the case of C$_2$H$_2$,  the characteristic radius in AA\,Tauri could not be determined 
and was set to that derived for HCN. Assuming that
the C$_2$H$_2$ emission is located at larger radii (e.\,g., $1\,\mathrm{AU}$ instead of $0.6\,\mathrm{AU}$), 
model MIX fits the observed column density, while model A predicts much lower values. Adversely, the spectrum of
model MIX shows C$_2$H$_2$ in weak absorption instead of emission.
From the previous discussion in Sections~\ref{sec_results_interaction_1} and~\ref{sec_results_interaction_2}, 
we can assume that stronger turbulence will transport
C$_2$H$_2$ higher up in the disk, leading to emission instead of absorption signatures.

The abundances of NH$_3$ are enhanced in model MIX, leading to absorption 
signatures in the spectrum between $8$ and $13\micron$.
To date, there has been no published detection of ammonia from protoplanetary disks,
but NH$_3$ has been found in \reftext{absorption} towards the hot core associated with
the massive protostar NGC\,7538\,IRS\,1 \citep{knez09}, using the Texas Echelon Cross Echelle Spectrograph (TEXES). 
Observing NH$_3$ emission from disks with high resolution and high sensitivity will allow us to draw important 
conclusions on the efficiency of turbulent mixing in protoplanetary disks.

Taking into account the model uncertainties, e.\,g. the H$_2$ formation rates on grains, 
the gas temperature profiles (which are related to the dust model adopted, e.\,g., to the dust growth and settling), 
as well as the uncertainties in deriving molecular column densities and characteristic radii
from the observations, further investigations are needed to strengthen the significance of our results.

In addition, the observed objects have inclination angles of $67^\circ$ for DR\,Tauri and $47^\circ$ for AS\,205A,
while our theoretical spectra are calculated for a disk seen face-on. This affects the observed line emission and the
derived column densities, since the chemical composition is different along the line of sight.
Robust conclusions therefore require spectral calculations accounting for the different inclination angles.

\section{Conclusion}\label{sec_conclusion}
We studied the chemical evolution of a steady protoplanetary disk, 
investigating different models for the formation rate of H$_2$ on dust grains and 
the effects of including viscous accretion, turbulent mixing and disk winds. 
Using our chemical results, we calculated the emergent near- and mid-infrared
spectrum from the disk, including contributions from the dust continuum and molecular line emission.

Our results show that the abundances of water and the hydroxyl radical are very sensitive to 
the H$_2$ formation rates on grains, with the resulting spectra differing considerably. 
The enhanced H$_2$ formation rates of \citet{cazaux02} lead to a drastic increase in the gas-phase
abundances of water and \reftext{OH} in the upper disk layers and \reftext{to strong}
emission lines. We concluded that dust grain settlement towards the midplane
and non-LTE calculations of the molecular line emission are required to obtain meaningful
results for models using these enhanced H$_2$ formation rates.

We investigated the effects of physical mass transport phenomena in the radial direction 
by viscous accretion and in the vertical direction
by diffusive turbulent mixing and disk winds. Due to the steady state assumption of the underlying protoplanetary disk model,
the wind velocities are limited to impede the evaporation of the disk over its lifetime 
and consequently, the effects of the disk winds on the chemistry are small. On the contrary, vertical diffusive
mixing has a considerable impact on the disk chemistry and increases the abundance 
of many observed species in the warm molecular
layer of the disk. Species which are produced efficiently in a particular layer and which are stable in lower (or upper) layers
can experience a significant enhancement due to the diffusive motion. Including the radial accretion flow
in the chemical network calculation leads to large changes in the chemical composition of the disk, 
particularly in the cold and dense midplane layer. 
The effect of the accretion flow on a particular species depends on the accretion timescale and the formation
and destruction timescales of that species. For example, inside $10\,\mathrm{AU}$, the gas-phase water abundances
are not overly affected by the accretion flow, while ammonia is greatly enhanced and methanol\reftext{,} greatly depleted.

Since the accretion flow affects mainly the midplane layers, 
the infrared molecular line emission is basically unchanged. Contrarily,
turbulent mixing alters the chemical composition of the warm molecular layer, where the molecular line emission is generated.
We thus contrasted two models with and without turbulent mixing to observations of protoplanetary disks around classical
T\,Tauri stars. We compared the modeled disk spectra and column densities of the upper layers of the disk to the observed
spectra and the derived column densities.
\pagebreak

In summary, a model with turbulent mixing reproduces the 
observations of AA\,Tauri, DR\,Tauri and AS\,205 better than a model
without mixing, although the discussion of the column 
densities in the upper layers and of the spectra suggests a more efficient vertical mass transport 
mechanism. In the models considered here, turbulent mixing provides an efficient
way of transporting material from the midplane into 
the transition layer. Due to the constant mixing velocity in each vertical
column, the turbulent timescales become longer than the chemical 
timescales at larger distances from the midplane. But these are the layers where
the disk winds studied by \citet{suzuki09} become most effective 
(the transition layer coincides roughly with $1.5$ disk scale heights)
and drive large-scale channel flows. Hence, a combination of 
turbulence in the cold midplane layers and disk winds in the transition layer,
both generated by the magneto-rotational instability, might account for 
the observations better than turbulent mixing or disk winds alone.
In the latter case, the steady state assumption would have to be abandoned 
to allow for stronger disk winds.

An interesting aspect for future studies will be the combination of radial 
and vertical transport processes. In this work, we showed that
the radial accretion flow alters predominantly the midplane 
abundances, while turbulent mixing affects the midplane and warm molecular
layers. We therefore expect that a combination of radial and vertical 
motion will have a strong impact on the chemical abundances in
the molecular layer, which will be reflected in the emission line spectrum.

\acknowledgments
The authors would like to thank Dr.\ Inutsuka and Dr.\ Suzuki from Nagoya University for 
providing the data of their MHD simulations and for
useful discussions. DH and HN were supported by the Grant-in-Aid for the Global 
COE Program ``The Next Generation of Physics, Spun from
Universality and Emergence'' from the Ministry of Education, Culture, Sports, 
Science and Technology (MEXT) of Japan. DH received further
support from the Kyoto University Startup Grants, and HN from the 
Grant-in-Aid for Scientific Research 2174~0137, MEXT.
CW acknowledges DEL for a studentship and JSPS for the award of a short-term fellowship to 
conduct research in Japan. Astrophysics at QUB is supported by a grant from the STFC.

The numerical calculations were carried out on Altix3700 BX2 at the Yukawa Institute for Theoretical Physics, Kyoto University,
and on the Cray XT4C at the Center for Computational Astrophysics, National Astronomical Observatory of Japan.
\clearpage

\appendix
\section{A. Physicochemical interactions -- numerics}\label{app_physicochemical_interaction} 
Modeling the physical mass transport phenomena in combination with calculating the chemical evolution 
is the key aspect of this study. 
Here, we discuss in detail the numerical techniques used to compute the chemistry accounting for transport mechanisms in
relation to the grid of our protoplanetary disk model.
\begin{figure*}
\centerline{\myincludegraphics{width=0.95\textwidth}{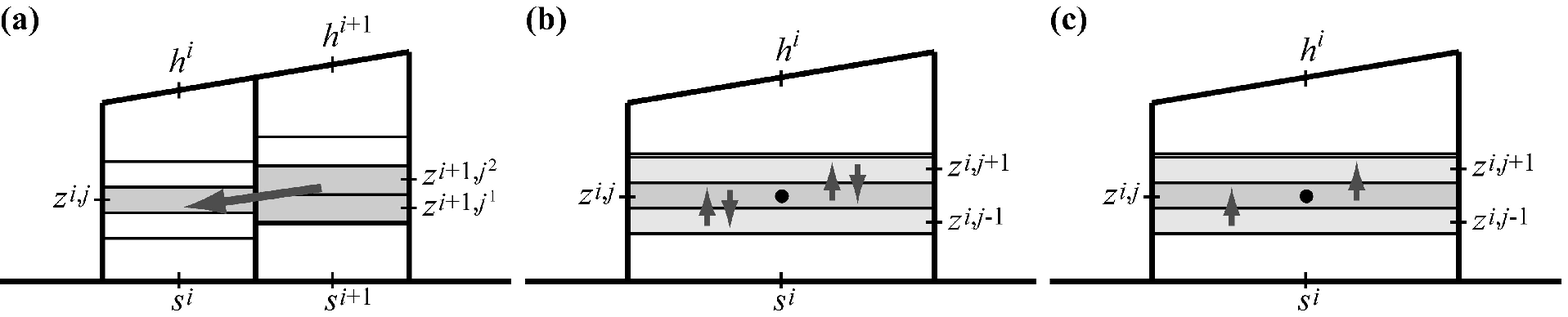}}
\caption{Schematic view of (a) accretion, (b) mixing and (c) wind modeling in {our model}.\label{fig_accretion_mixing_wind}}
\end{figure*}

The general expression for the rate of change in the abundance of
gas-phase species $k$ in grid cell $(i,j)$ at radius $s^i$ and vertical distance $z^j$ from the midplane is given by
\begin{equation}
\dot{n}_k^{i,j} = \frac{\Delta n_k^{i,j}}{\Delta t} = P_k^{i,j} - L_k^{i,j} - \dot{n}_{k\,\mathrm{ice}}^{i,j} + \dot{n}_{k,\,\mathrm{acc/mix/wind}}^{i,j}\,,
\end{equation}
where $P_k$ and $L_k$ denote the production and loss terms due to gas-phase chemical reactions, 
$\dot{n}_{k\,\mathrm{ice}}$ the adsorption/desorption rate ($>0$ for net freeze-out,
$<0$ for net {evaporation}) and $\dot{n}_{k,\,\mathrm{acc/mix/wind}}$ the rates arising 
from mass transport due to accretion, mixing or disk winds.
In the case of molecular hydrogen, an additional (positive) term 
$\dot{n}_{\mathrm{H}_2}$ is added to the right-hand side to account for the
H$_2$ formation on grains. Accordingly, in the case of atomic hydrogen, an 
additional term $-2 \dot{n}_{\mathrm{H}_2}$ is required.

\subsection{A.1 Radial transport by viscous accretion}\label{app_accretion} 
We assume that inward accretion occurs along constant streamlines $z/h$.
(see Figure~\ref{fig_accretion_mixing_wind}(a) for an illustration of the coupling between the grid cells at different radii).
The net accretion rate along constant streamlines is given in 
Equation~\eqref{eqn_accretion_rate}, where the differentiation with respect to $l$
needs to be translated into a differentiation with respect to $s$. 
Accordingly, on the discrete grid of the disk model, the net accretion rate for
cell $(i,j)$ can be expressed as 
\begin{equation}
\dot{n}_{k,\,\mathrm{acc}}^{i,j} = 	\frac{\bar{v}_s^{i,i+1}}{\Delta s^{i}} \left\{ - n_k^{i,j}
		+ \frac{\tilde{\rho}^{i+1,j^1\!,j^2\!}}{\rho^{i,j}} \tilde{n}_k^{i+1,j^1\!,j^2\!} \right\}
\label{eqn_accretion_rate_discrete}
\end{equation}
where $\tilde{\rho}$ and $\tilde{n}$ represent the average mass and number density of the
 outer grid cells, weighed by their relative mass transfer into
cell $(i,j)$. The accretion velocity $\bar{v}_s^{i,i+1}$ is evaluated at the 
interface between the grid cells $i$ and $i+1$ and is
introduced to improve the accuracy of the finite calculation. 
The additional factor $\tilde{\rho}^{i+1,j^1\!,j^2\!}/\rho^{i,j}$ arises from the requirement of mass conservation.
In the context of a steady disk model, all variables except $n_k^{i,j}$ and $\tilde{n}_k^{i+1,j^1\!,j^2}$ are constant 
with time.

\subsection{A.2 Vertical transport by turbulent mixing}\label{app_mixing} 
Figure~\ref{fig_accretion_mixing_wind}(b) demonstrates the grid-cell mass transfer by turbulent mixing in the diffusion-type model.
Starting from Equation~\eqref{eqn_mixing_rate}, we replace the differentials by finite differences over the vertical 
extent $\Delta z$ of the grid cells\reftext{.}
Additionally, we introduce mean densities $\bar{\rho}$ and velocities $\bar{v}$ at the interface of the grid cells as before.
The exchange rates of species $k$ between grid cell $(i,j)$ and the adjacent grid cells $(i,j-1)$ and $(i,j+1)$ then become
\begin{equation}
\dot{n}_{k,\,\mathrm{mix}}^{i,j} =
	\frac{\bar{\rho}^{i,j-1,j}}{\rho^{i,j-1}} \cdot \frac{\bar{v}_z^{i,j-1,j}}{\Delta z^{i,j}} \cdot 
		\left(n_k^{i,j-1} - \frac{\rho^{i,j-1}}{\rho^{i,j}}\,n_k^{i,j} \right)
	+\frac{\bar{\rho}^{i,j,j+1}}{\rho^{i,j}} \cdot \frac{\bar{v}_z^{i,j,j+1}}{\Delta z^{i,j}} \cdot 
		\left(\frac{\rho^{i,j}}{\rho^{i,j+1}}\,n_k^{i,j+1} - n_k^{i,j} \right).\label{eqn_mixing_rate_discrete}
\end{equation}

\subsection{A.3 Vertical transport by disk winds}\label{app_wind} 
The model of the upward disk wind is shown in Figure~\ref{fig_accretion_mixing_wind}(c).
In the finite case, the differential equation for the wind rate (Equation~\eqref{eqn_wind_rate}) of grid cell $(i,j)$ translates into
\begin{equation}
\dot{n}_{k,\,\mathrm{wind}}^{i,j} = \frac{v_z^{i,j-1,j}}{{\Delta z}^{i,j}} \cdot n_k^{i,j-1} 
- \frac{v_z^{i,j,j+1}}{{\Delta z}^{i,j}} n_k^{i,j}\,.\label{eqn_wind_rate_discrete}
\end{equation}
In both the wind and the mixing scenarios, the only variable changing with time is $n_k$.
\pagebreak

\end{document}